\documentclass[aps,prd,twocolumn,groupedaddress,10pt]{revtex4-2}
\usepackage{graphicx}
\usepackage{dcolumn}
\usepackage{bm}
\usepackage{amsmath} 
\usepackage{amssymb}
\usepackage{xcolor}
\usepackage{hyperref}
\hypersetup{
	colorlinks=true,
	linkcolor=red,
	filecolor=magenta,      
	urlcolor=blue,
	pdftitle={Selective Thermalization},
	pdfpagemode=FullScreen,
	citecolor= blue,
}
\begin{document}



\title{Field-Dependent Particle Excitations in Null-Shifted Rindler Wedges}


%
\author{Rakesh K Jha}
	\email{jhark007.qg@gmail.com}

	\affiliation{Department of Physics, Birla Institute of Technology and Science, Pilani-Hyderabad Campus \\Hyderabad, 500078, India}	
	\date{\today}


\date{\today}

\begin{abstract}
The Unruh effect establishes that the Minkowski vacuum, when restricted to a Rindler wedge and described in terms of boost eigenmodes, satisfies the Kubo-Martin-Schwinger (KMS) condition and is therefore perceived as a thermal state by uniformly accelerated observers. This result is robust and remains valid for both massless and massive fields. In this work, we investigate a distinct geometric setup involving two null-shifted, nested Rindler wedges $R_2 \subset R_1$, where the standard assumptions underlying the Unruh effect are not directly applicable. We analyse how the Rindler vacuum defined with respect to $R_1$ is represented in terms of mode functions adapted to $R_2$ by computing overlap coefficients defined through projection of field modes along null directions. Owing to the absence of a common global Cauchy surface for the nested wedge pair, these coefficients are defined in a formal sense and do not correspond to a unitary Bogoliubov transformation between complete Fock spaces. Our results show that the projected mode content exhibits nontrivial mixing, but this does not constitute a violation of the KMS condition nor a modification of the standard Unruh effect. Instead, it reflects the limitations of defining particle content through partial, null-based projections in a restricted geometric setting. The analysis is carried out for both scalar and Dirac fields, with particular attention to the role of mode structure, inner products, and consistency conditions.

\vspace{0.5cm}
\end{abstract}


\maketitle{}

\section{Introduction\label{Sec-1}}

The Unruh effect provides a fundamental link between quantum field theory and observer-dependent notions of particles\cite{Unruh:1976db, Hawking:1975vcx, Hawking1975}. It establishes that the Minkowski vacuum, when restricted to a Rindler wedge and expressed in terms of boost eigenmodes, satisfies the Kubo–Martin–Schwinger (KMS)\cite{Haag:1992hx} condition and is therefore perceived as a thermal state by uniformly accelerated observers. This result is robust and applies to both massless\cite{Jha:2025tpg, Jha:2026rmu} and massive quantum fields\cite{Crispino:2007eb}; while the presence of mass modifies the detailed structure of the mode functions and detector response, it does not alter the underlying thermal character associated with boost evolution.

In this work, we investigate a geometrically distinct setting that departs from the standard Unruh construction. Rather than comparing Minkowski and Rindler quantisations, we consider two null-shifted, nested Rindler wedges\cite{Lochan:2021pio, Kolekar:2013hra}, denoted by $R_2 \subset R_1$, and analyse the relation between the corresponding mode decompositions. Specifically, we take the vacuum state\cite{Aalsma:2019rpt, Sriramkumar:1999nw} defined with respect to the Rindler modes in $R_1$ and examine its representation in terms of modes adapted to the subwedge $R_2$.

This configuration differs from the standard framework in several important respects. First, both mode systems are associated with non-inertial observers, and the comparison does not involve the Minkowski vacuum. Second, the null translation relating the two wedges is not generated by a spacetime isometry that preserves positive frequency\cite{Padmanabhan:2010zzb}. Consequently, the relation between the two sets of modes does not fall within the usual framework of Bogoliubov transformations between complete orthonormal bases defined on a common global Cauchy surface\cite{Wald1993}.

A further structural feature of the setup is that the wedges are nested, with $R_2$ contained within $R_1$, and the corresponding mode functions have support in different spacetime regions. As a result, the two mode decompositions do not, in general, define complete bases of a common global solution space. In particular, there is no single Cauchy surface\cite{Wald:1984rg} on which both sets of modes simultaneously provide complete initial data. To address this issue, we restrict the analysis to the overlap region where both coordinate systems are well defined and construct the relevant inner products within this domain. The resulting coefficients should therefore be understood as projection-based quantities\cite{Mukhanov:2007zz}, rather than as elements of a unitary transformation between Fock spaces.

Within this framework, we compute the corresponding mode overlap coefficients for both massive scalar and Dirac fields. In the scalar case, the presence of mass introduces nontrivial radial dependence in the mode functions, typically expressed in terms of modified Bessel functions\cite{Crispino:2007eb}. In the fermionic case, the analysis requires the use of the Dirac inner product and appropriate spinor solutions\cite{Jha:2025tpg}. This allows for a consistent comparison of mode mixing in both bosonic and fermionic sectors.

Our aim is not to establish thermality, but rather to analyse how mode decompositions and particle interpretations depend on the geometric relation between non-inertial coordinate systems. The mixing between positive- and negative-frequency components that arises in this context reflects the limitations of defining particle content through restricted, null-based projections, and does not imply any modification of the standard Unruh effect or the associated KMS condition\cite{Kia_ka_2018}.

From this perspective, the mode mixing obtained in the null-shifted wedge configuration can be interpreted as a form of selective particle excitation\cite{Jha:2025tpg} arising from geometric projection rather than genuine thermality. Since the coefficients are defined through overlaps restricted to a subregion and are not tied to a global notion of positive frequency, the resulting particle content depends sensitively on the chosen null sector and on the structure of the mode functions. This leads to a non-universal distribution of excitations, in contrast to the Planckian spectrum characteristic of the Unruh effect\cite{Unruh:1976db, Unruh:1983ms, UtiyamaDeWitt1962}.

It is useful to situate the present analysis within the broader conceptual framework of thermal phenomena in quantum field theory. In standard settings, thermality associated with horizons can be understood through several interrelated structures: (i) restriction of a global vacuum state to a subregion, leading to a mixed state; (ii) modular (Tomita–Takesaki) flow and its relation to geometric symmetries\cite{Longo:2017bae}; (iii) the Kubo–Martin–Schwinger (KMS) condition characterising equilibrium states; and (iv) detector-based responses probing field correlations along specific worldlines. In the Unruh effect, these elements are mutually consistent and ensure that the Minkowski vacuum exhibits a universal thermal character with respect to boost evolution\cite{Earman:2011zz}. In contrast, the present construction does not realise this full structure, as it involves a comparison between mode decompositions associated with geometrically distinct, nested wedges rather than the restriction of a global state to a single causal domain.

The massless limit is briefly discussed in the conclusion, where its connection to conformal field theory becomes manifest.


\section{Geometric SetUp and Quantum Field Theoretic Framework \label{Sec-2}}

\subsection{Rindler coordinates \label{Subsec-2.1}}

We work in $(1+1)$-dimensional Minkowski spacetime with inertial coordinates $(T, X)$ and introduce Rindler coordinates adapted to uniformly accelerated observers \cite{Rindler:1960zz}. We consider two Rindler wedges, denoted by $R_1$ and $R_2$, described respectively by coordinate systems $(t_1,x_1)$ and $(t_2,x_2)$.

As illustrated in Fig.~\ref{Fig:1}, the wedge $R_2$ lies entirely within $R_1$, so that $R_2 \subset R_1 \subset M$.
The boundaries of these wedges are null lines in Minkowski spacetime. The essential geometric feature is that $R_2$ is obtained from $R_1$ through a translation along a null direction, as shown in Figs.~\ref{Fig:1} and \ref{Fig:2}. These null shifts preserve the causal structure of the wedges while modifying the relation between their associated null coordinates.

\subsubsection{Minkowski to Rindler spacetime}

For the wedge $R_1$, the transformation between Minkowski and Rindler coordinates is given by
\begin{align}
T &= \frac{e^{a x_1}}{a} \sinh(a t_1), \label{Eq:2.1.0.1} \\
X &= \frac{e^{a x_1}}{a} \cosh(a t_1). \label{Eq:2.1.0.2}
\end{align}

Introducing null coordinates
$u_1 = t_1 - x_1, \qquad v_1 = t_1 + x_1,$

the corresponding Minkowski null coordinates become
\begin{align}
U_M &= T - X = -\frac{e^{-a u_1}}{a}, \label{Eq:2.1.0.3} \\
V_M &= T + X = \frac{e^{a v_1}}{a}. \label{Eq:2.1.0.4}
\end{align}

Here $a$ denotes the proper acceleration. Since different values of $a$ correspond only to different parametrisations of the same wedge, we take $a_1 = a_2 = a$ for simplicity.

\subsubsection{Null shifts along the \texorpdfstring{$V$}{V}-direction}

\begin{figure}[!ht]
\centering
\includegraphics[scale=0.7]{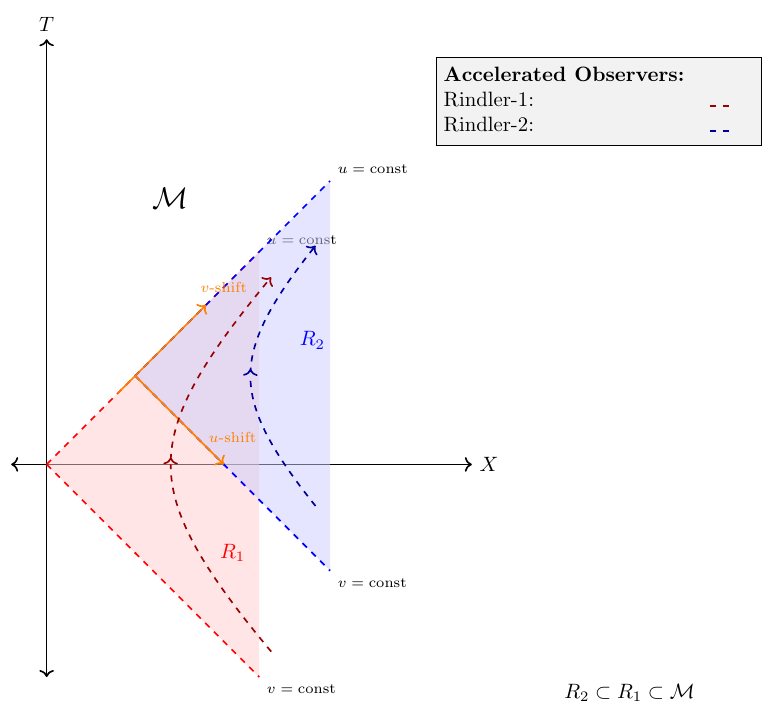}
\caption{Null-shifted Rindler wedges $R_1$ and $R_2$ along the $V$-direction.\label{Fig:1}}
\end{figure}

We now construct the wedge $R_2$ via a translation along the $V$-direction, as illustrated in Fig.~\ref{Fig:1}.

The transformation to $(t_2,x_2)$ is given by,
\begin{align}
T &= \frac{e^{a x_2}}{a} \sinh(a t_2) + \Delta_2,  \label{Eq:2.1.0.5}\\
X &= \frac{e^{a x_2}}{a} \cosh(a t_2) + \Delta_2. \label{Eq:2.1.0.6}
\end{align}

This corresponds to a constant shift in the null coordinate $V_M$, leading to
\begin{align}
U_M &= -\frac{e^{-a u_2}}{a},  \label{Eq:2.1.0.7}\\
V_M &= \frac{e^{a v_2}}{a} + 2\Delta_2.  \label{Eq:2.1.0.8}
\end{align}

Comparing  Eqs.~(\ref{Eq:2.1.0.3}) and (\ref{Eq:2.1.0.4}) with  Eqs.~(\ref{Eq:2.1.0.7}) and (\ref{Eq:2.1.0.8}) , we obtain
\begin{align}
u_1 &= u_2,  \label{Eq:2.1.0.9}\\
\frac{e^{a v_1}}{a} &= \frac{e^{a v_2}}{a} + 2\Delta_2. \label{Eq:2.1.0.10}
\end{align}

Thus,
\begin{equation*}
v_1 = \frac{1}{a} \ln\left(e^{a v_2} + 2a\Delta_2 \right).
\end{equation*}

The parameter $\Delta_2$ characterises the magnitude of the null displacement. It enters the transformation through the dimensionless combination $a\Delta_2$, which encodes the relative separation between the two wedges. In general, different values of $a\Delta_2$ correspond to physically distinct configurations.

For the present analysis, we choose a specific value
\begin{equation}
\Delta_2 = \frac{1}{2a}, \label{Eq:2.1.0.11}
\end{equation}
which simplifies the coordinate relation to
\begin{equation}
v_1 = \frac{1}{a} \ln\left(e^{a v_2} + 1 \right), \label{Eq:2.1.0.12}
\end{equation}
This choice is made for calculational convenience and does not affect the qualitative structure of the transformation. The general case can be recovered straightforwardly by restoring the dependence on $a\Delta_2$.

\subsubsection{Null shifts along the \texorpdfstring{$U$}{U}-direction}

\begin{figure}[!ht]
\centering
\includegraphics[scale=0.7]{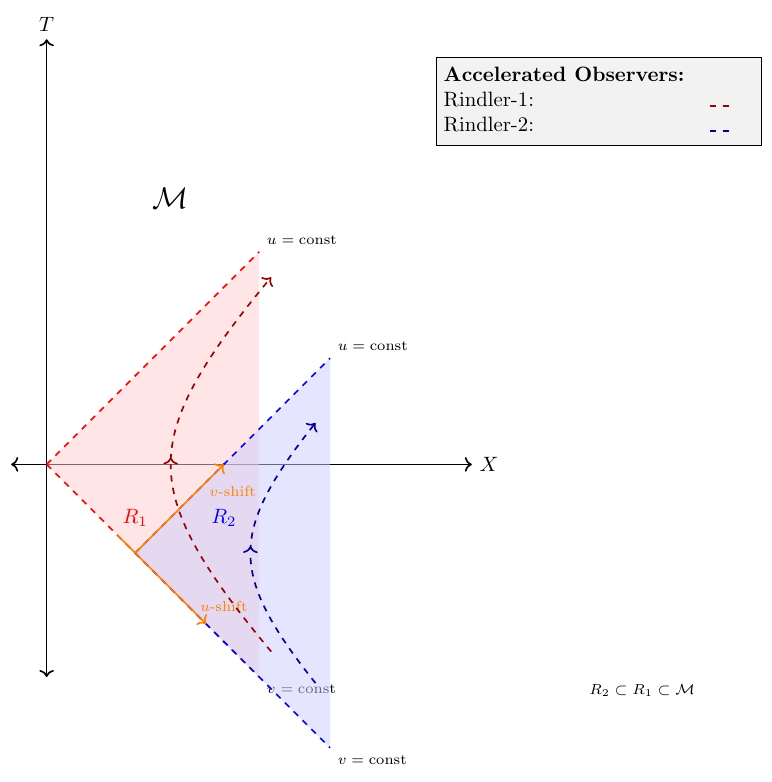}
\caption{Null-shifted Rindler wedges $R_1$ and $R_2$ along the $U$-direction.\label{Fig:2}}
\end{figure}

We now consider shifts along the $U$-direction, as shown in Fig.~\ref{Fig:2}.

The transformation is given by,
\begin{align}
T &= \frac{e^{a x_2}}{a} \sinh(a t_2) - \Delta_2, \label{Eq:2.1.0.13} \\
X &= \frac{e^{a x_2}}{a} \cosh(a t_2) + \Delta_2. \label{Eq:2.1.0.14}
\end{align}
leading to,
\begin{align}
U_M &= -\frac{e^{-a u_2}}{a} - 2\Delta_2, \label{Eq:2.1.0.15} \\
V_M &= \frac{e^{a v_2}}{a}. \label{Eq:2.1.0.16}
\end{align}

Comparing with the $R_1$ expressions, we obtain
\begin{align}
v_1 &= v_2, \label{Eq:2.1.0.17}\\
u_1 &= -\frac{1}{a} \ln\left(e^{-a u_2} + 2a\Delta_2 \right). \label{Eq:2.1.0.18}
\end{align}

With the same choice $\Delta_2 = 1/(2a)$, this reduces to
\begin{equation}
u_1 = -\frac{1}{a} \ln\left(e^{-a u_2} + 1 \right), \label{Eq:2.1.0.19}
\end{equation}
These relations define the mapping between the two nested Rindler wedges generated by null translations along the $U$ and $V$ directions.

\subsection{Quantum field setup, mode decomposition and Bogoliubov relations \label{Subsec-2.2}}

We consider a real scalar field $\hat{\Phi}$ propagating on $(1+1)$-dimensional Minkowski spacetime $\mathcal{M}$. The field satisfies the Klein-Gordon equation and admits a standard quantisation in terms of globally defined mode functions on $\mathcal{M}$.

The quantisation may be formulated on a global Cauchy surface $\Sigma \subset \mathcal{M}$ (for example $T=\mathrm{const}$), with the Klein-Gordon inner product\cite{Wald1993}
\begin{equation}
(\phi_1,\phi_2)= -i \int_{\Sigma} d\Sigma^\mu \left(\phi_1 \partial_\mu \phi_2^* - \phi_2^* \partial_\mu \phi_1 \right).
\label{Eq:2.2.0.1}
\end{equation}
A complete orthonormal set of positive-frequency solutions with respect to the global timelike Killing vector $\partial_T$ is denoted by $\{u_k^{(M)}\}$, and the field admits the expansion
\begin{equation}
\hat{\Phi} = \int_0^\infty dk \left( \hat{a}_k^{(M)} u_k^{(M)} + \hat{a}_k^{(M)\dagger} u_k^{(M)*} \right),
\label{Eq:2.2.0.2}
\end{equation}
With canonical commutation relations
\begin{equation}
[\hat{a}_k^{(M)},\hat{a}_{k'}^{(M)\dagger}] = \delta(k-k').
\label{Eq:2.2.0.3}
\end{equation}
The Minkowski vacuum $|0_M\rangle$ is defined by $\hat{a}_k^{(M)} |0_M\rangle = 0$.

In addition to this global description, one may introduce Rindler coordinates adapted to uniformly accelerated observers confined to a wedge $R_i$, with time evolution generated by the boost Killing vector\cite{BirrellDavies, Fabbri:2005mw} $\partial_{t_i}$. This leads to a corresponding set of mode functions $\{f_i(\omega)\}$, defined as positive-frequency solutions with respect\cite{Padmanabhan:2010xe} to $\partial_{t_i}$ and supported within $R_i$.

However, these Rindler modes are not, in general, smooth global solutions on $\mathcal{M}$ and do not provide complete initial data on the global Cauchy surface $\Sigma$. While each wedge $R_i$ is itself globally hyperbolic and admits its own Cauchy surfaces\cite{Frodden:2018mdm}, these surfaces are not Cauchy surfaces for the full spacetime $\mathcal{M}$. As a result, the mode sets $\{f_1\}$ and $\{f_2\}$ associated with two wedges $R_1$ and $R_2$ do not constitute complete orthonormal bases of a single global solution space.

In the present setup, the wedges are related by a null translation and are nested such that $R_2 \subset R_1$. The corresponding mode functions are therefore supported on different spacetime regions. In particular, there is no global Cauchy surface on which both sets of modes provide complete initial data. Consequently, the standard framework of Bogoliubov transformations between complete bases on a common Cauchy surface does not directly apply.

To compare the mode decompositions, one must therefore restrict attention to the spacetime region where both coordinate systems are well-defined. Any relations between the mode sets $\{f_1\}$ and $\{f_2\}$ are to be understood as formal expansions within this overlap region, rather than as transformations between complete bases of the same Hilbert space.

In particular, an expression of the form
\begin{equation}
f_2(\Omega) = \int_0^\infty d\omega \left( \alpha_{21}(\Omega,\omega) f_1(\omega) + \beta_{21}(\Omega,\omega) f_1^*(\omega) \right)
\label{Eq:2.2.0.4}
\end{equation}
does not define a genuine Bogoliubov transformation in the standard sense. The coefficients $\alpha_{21}$ and $\beta_{21}$ are not guaranteed to satisfy the canonical Bogoliubov identities, and no unitary map between the corresponding Fock spaces is implied.

Accordingly, quantities constructed from such coefficients, for example
\begin{equation}
\langle 0_{R_1} | \hat{a}_2^\dagger(\Omega)\hat{a}_2(\Omega) | 0_{R_1} \rangle 
= \int_0^\infty d\omega \, |\beta_{21}(\Omega,\omega)|^2,
\label{Eq:2.2.0.5}
\end{equation}
should be interpreted only as measures of the mismatch between the two mode decompositions. They do not represent particle production in the standard sense, which requires a unitary Bogoliubov transformation between complete mode bases defined on a common Cauchy surface.

\subsection{Vacuum choice, restriction to subregions and thermality \label{Subsec-2.3}}

The distinction between Minkowski and Rindler quantisations extends to the corresponding vacuum states\cite{Aalsma:2019rpt, Sriramkumar:1999nw,SRIRAMKUMAR_2002}. The Minkowski vacuum $|0_M\rangle$ is globally defined and regular across $\mathcal{M}$, whereas the Rindler vacuum $|0_{R_i}\rangle$, defined by $\hat{a}_i(\omega)|0_{R_i}\rangle = 0$, is only defined relative to observables supported in the wedge $R_i$ and is singular at the corresponding horizon.

In the present situation, where $R_2 \subset R_1$, the algebra of observables satisfies $\mathcal{A}(R_2) \subset \mathcal{A}(R_1)$. The physically relevant state in $R_2$ obtained from the $R_1$ vacuum is therefore given by restriction,
\begin{equation}
\omega_{R_2}(\mathcal{O}) = \langle 0_{R_1} | \mathcal{O} | 0_{R_1} \rangle,  \label{Eq:2.3.0.1}
\qquad \mathcal{O} \in \mathcal{A}(R_2).
\end{equation}

This restricted state is, in general, mixed, reflecting the fact that the vacuum $|0_{R_1}\rangle$ is not pure when viewed as a state on the subalgebra $\mathcal{A}(R_2)$. It does not coincide with the vacuum defined by the $R_2$ mode decomposition.

It is important to separate the notion of particle content from thermality. A non-zero expectation value of a number operator defined with respect to a particular mode decomposition reflects a mismatch between mode definitions and does not, by itself, imply the presence of a thermal state. Thermality is instead characterised by the Kubo-Martin-Schwinger (KMS) condition on correlation functions with respect to a specified time evolution.

In particular, even if the expectation value of a number operator\cite{Birrell:1982ix, Mukhanov:2007zz,fabbri12005} takes a Planckian form,
\begin{equation}
\langle \hat{N}(\Omega)\rangle \sim \frac{1}{e^{\beta \Omega}-1}, \label{Eq:2.3.0.2}
\end{equation}

More precisely, the KMS condition must be defined with respect to a specified one-parameter automorphism group of the algebra, typically generated by the relevant time evolution (e.g. the boost generator in a Rindler wedge). In the absence of such a well-defined modular flow associated with the state $\omega_{R_2}$, a Planckian distribution cannot be taken as evidence of thermality.

\section{Thermal properties of massive Scalar fields in null-shifted Rindler wedges \label{Sec-3}}
In this section, we analyse the behaviour of a massive scalar field in $(1+1)$-dimensional Minkowski spacetime, restricted to the wedge $\mathrm{R}_2$ obtained from $\mathrm{R}_1$ by a translation along one of the null directions, either the $U$ or $V$-axis, in light-cone coordinates. This framework provides a natural generalisation of our earlier results on selective particle excitations in Rindler spacetime~\cite{Jha:2025tpg}.
\subsection{The Field Equation\label{Subsec-3.1}}
We begin with a real scalar field $\phi(t,x)$ of mass $m$, obeying the Klein-Gordon equation
\begin{equation}
 \nabla^\mu \;\nabla_\mu \phi(t,x) - m^2\;\phi(t,x)   =0,\label{Eq:3.1.0.1}
\end{equation}
In component form, this becomes
\begin{equation}
   g^{\mu\nu}\; \nabla_\mu\;\nabla_\nu \phi(t,x) -  m^2\;\phi(t,x)   =0,\label{Eq:3.1.0.2}
\end{equation}
The line element in $(1+1)$-dimensional Rindler coordinates takes the conformally flat form,
\begin{equation}
    ds^2 = e^{2\;a\;x}\;(dt^2 - dx^2),\label{Eq:3.1.0.3}
\end{equation}
From which the metric and its inverse are given by,
\begin{align}
g_{\mu\nu} =
  \begin{pmatrix}
    e^{\;2\; a\; x} && 0\\
    0  && - e^{\;2\;a\;x}
 \end{pmatrix} ;
g^{\mu\nu} =
  \begin{pmatrix}
    e^{\;-2 \;a \;x} && 0\\
    0  && - e^{\;-2\;a\;x}
 \end{pmatrix}, \label{Eq:3.1.0.4}
 \end{align}
 Substituting into Eq.~\eqref{Eq:3.1.0.2}, the Klein-Gordon equation becomes
 \begin{equation}
  e^{\;-2\;a\;x}\; \frac{\partial^2\phi (t,x)}{\partial t^2} - e^{\;-2\;a\;x}\; \frac{\partial^2\phi(t,x)}{\partial x^2} - m^2\;\phi(t,x) =0 ,\label{Eq:3.1.0.5}
 \end{equation}
 Which simplifies to,
\begin{equation}
   \bigg(\frac{\partial^2}{\partial t^2}
   - \frac{\partial^2}{\partial x^2}
   - m^2 e^{2 a x}\bigg)\phi(t,x) = 0 , \label{Eq:3.1.0.6}   
\end{equation}

We adopt the separable ansatz,
\begin{equation}
   \phi(t,x) = e^{-i \omega t}\,\psi(x),  \label{Eq:3.1.0.7}  
\end{equation}

Substituting Eq.~\eqref{Eq:3.1.0.7}into Eq.~\eqref{Eq:3.1.0.6} and using 
$\partial_t^2 \phi = -\omega^2 \phi$, we obtain
\begin{equation}
   \frac{d^2 \psi}{dx^2}
   + \bigg(\omega^2 - m^2 e^{2 a x}\bigg)\psi(x) = 0,  \label{Eq:3.1.0.8}  
\end{equation}

To solve this equation, we introduce the variable
\begin{equation}
\begin{split}
   z &= \frac{m}{a} e^{a x}, \\
   \frac{d}{dx} &= a z \frac{d}{dz}, \\
   \frac{d^2}{dx^2} &= a^2 z^2 \frac{d^2}{dz^2}
   + a^2 z \frac{d}{dz}.  \label{Eq:3.1.0.9}
\end{split}
\end{equation}

Substituting the above expressions into Eq.~\eqref{Eq:3.1.0.8}, we find
\begin{equation}
   z^2 \frac{d^2 \psi}{dz^2}
   + z \frac{d \psi}{dz}
   + \bigg(\frac{\omega^2}{a^2} - z^2\bigg)\psi(z) = 0,  \label{Eq:3.1.0.10}  
\end{equation}

  This is the standard form of the modified Bessel equation~\cite{olver10},
\begin{equation*}
   z^2 \psi'' + z \psi' - (z^2 + \nu^2)\psi = 0
\end{equation*}
With,
\begin{equation}
   \nu^2 = -\frac{\omega^2}{a^2}
   \quad \Rightarrow \quad
   \nu = \pm i\,\frac{\omega}{a},  \label{Eq:3.1.0.11}
\end{equation} 
Hence, the decaying solutions are
\begin{equation}
\psi(x) = N_\omega \, K_{\frac{i\omega}{a}}\!\left(\frac{m}{a}e^{ax}\right), 
\qquad
\psi(x) = N_\omega \, K_{-\frac{i\omega}{a}}\!\left(\frac{m}{a}e^{ax}\right). \label{Eq:3.1.0.12}
\end{equation}

Using the connection formula, valid for all $\nu \in \mathbb{C}$ and $z>0$~\cite{olver10},
\begin{equation}
K_{-\nu}(z) = K_{\nu}(z),
\label{Eq:3.1.0.13}
\end{equation}
Although $K_{-\nu}=K_\nu$, the independent mode structure arises from time dependence and sector decomposition rather than the radial function alone.
We see that the radial dependence of the two solutions coincides. However, this identity does not imply that the corresponding physical mode functions are identical, since the full solutions differ in their time dependence and mode structure.

We work in conformally flat Rindler coordinates and decompose the field into null-directed mode sectors. A convenient set of mode functions is
\begin{equation}
u_\omega(t_1,x_1) = N_\omega \, e^{-i\omega t_1}\, 
K_{\frac{i\omega}{a}}\!\left(\frac{m}{a}e^{a x_1}\right),
\label{Eq:3.1.0.14}
\end{equation}
together with their complex conjugates.
These mode functions are defined with respect to the Rindler time $t_1$ and are supported within the wedge $R_1$. They do not constitute a complete set of global solutions on Minkowski spacetime, but rather define a wedge-adapted mode decomposition.

The full field operator expansion then takes the form
\begin{equation}
\begin{split}
\hat{\phi}(t_1,x_1) = \int_{0}^{\infty} d\omega \bigg[
&\overrightarrow{\hat a}(\omega)\, u_\omega(t_1,x_1)
+ \overrightarrow{\hat a}^\dagger(\omega)\, u_\omega^*(t_1,x_1) \\
&+ \overleftarrow{\hat b}(\omega)\, u_\omega(t_1,x_1)
+ \overleftarrow{\hat b}^\dagger(\omega)\, u_\omega^*(t_1,x_1)
\bigg],
\end{split}
\label{Eq:3.1.0.15}
\end{equation}

Although both sectors involve the same radial function due to the identity 
$K_{-\nu}(z)=K_\nu(z)$, the corresponding operators 
$\overrightarrow{\hat a}(\omega)$ and $\overleftarrow{\hat b}(\omega)$ 
remain independent. These operators encode distinct mode sectors of the 
field and must be retained separately to preserve a complete 
set of solutions.

Here, $N_\omega$ denotes the normalisation factor, which is determined using the Klein-Gordon inner product\cite{Frodden:2018mdm, Crispino:2007eb, Wald1993} in $(1+1)$ dimensions. In conformally flat coordinates, the Klein-Gordon inner product is invariant under Weyl rescaling, and therefore reduces to the flat-space expression. Consequently, constant $t_1$ hypersurfaces provide a valid Cauchy surface for normalisation.

The inner product between two modes is defined as\cite{wald1995quantumfieldtheorycurved},
\begin{equation}
\langle h(\omega)|h(\omega^\prime)\rangle 
= i \int_{-\infty}^{\infty} dx_1 
\left[ h^*(\omega^\prime)\,\partial_{t_1} h(\omega) 
- h(\omega)\,\partial_{t_1} h^*(\omega^\prime) \right].
\label{Eq:3.1.0.16}
\end{equation}

Substituting the mode functions defined in Eq.~(\ref{Eq:3.1.0.14}) into Eq.~(\ref{Eq:3.1.0.16}), one obtains
\begin{equation}
\begin{split}
\langle h(\omega)|h(\omega^\prime)\rangle 
= \frac{N_{\omega^\prime} N_{\omega} (\omega+\omega^\prime)}{a} 
\, e^{i(\omega^\prime - \omega)t_1} \\
\int_{0}^{\infty} \frac{dp}{p} 
\, K_{\frac{i\omega^\prime}{a}}(p)\, K_{\frac{i\omega}{a}}(p).
\end{split}
\label{Eq:3.1.0.17}
\end{equation}

To evaluate the integral in Eq.~(\ref{Eq:3.1.0.17}), we use the standard orthogonality relation for the modified Bessel (Macdonald) functions~\cite{Gradshteyn:1943cpj,olver1997asymptotics,olver10}:
\begin{equation}
\int_{0}^{\infty} \frac{dx}{x} 
\, K_{i\nu}(x)\, K_{i\nu'}(x) 
= \frac{\pi^2}{2\,\nu\,\sinh(\pi \nu)}\, \delta(\nu - \nu'), \label{Eq:3.1.0.18}
\end{equation}

Using Eqs.~(\ref{Eq:3.1.0.17}) and (\ref{Eq:3.1.0.18}), and imposing the normalization condition
$
\langle h(\omega)|h(\omega^\prime)\rangle = \delta(\omega - \omega^\prime),
$
We obtain the normalisation coefficient as
\begin{equation}
N_\omega = \frac{1}{\pi} 
\left( \frac{1}{a}\,\sinh\!\left(\frac{\pi \omega}{a}\right) \right)^{1/2}.
\label{Eq:3.1.0.19}
\end{equation}
Thus, the normalised mode expansion becomes:
\begin{equation}
    \begin{split}
         \hat{\phi}(x_1,t_1) = \int_{0}^{\infty}d\omega\; \frac{1}{\pi}\;\bigg(\frac{1}{a}\;sinh\;(\frac{\pi\;\omega}{a})\bigg)^{\frac{1}{2}} K_{\;\frac{i\;\omega}{a}} \bigg(\frac{m}{a} e^{\;a\;x_1}\bigg)\\
         \bigg[e^{\;- i\;\omega\; t_1}\bigg ( \overrightarrow{\hat{a}}(\omega)+\overleftarrow{\hat{b}}(\omega)\bigg )+ e^{\;i\;\omega \;t_1}\bigg(\overrightarrow{\hat{a}^\dagger}(\omega)+\overleftarrow{\hat{b}^\dagger}(\omega)\bigg) \bigg],\label{Eq:3.1.0.20}
    \end{split}
\end{equation}
To analyse the near-horizon behaviour of the field, we consider the regime 
$z \to 0$~\cite{Crispino:2007eb}, corresponding to $x_1 \to -\infty$, and use the small-argument asymptotic expansion of the modified Bessel function~\cite{Gradshteyn:1943cpj},
\begin{equation}
K_{\nu}(z) \sim \frac{\pi}{2\sin(\pi \nu)}\left[
\frac{(\frac{1}{2}z)^{-\nu}}{\Gamma(1-\nu)} 
- 
\frac{(\frac{1}{2}z)^{\nu}}{\Gamma(1+\nu)}
\right], \label{Eq:3.1.0.21}
\end{equation}

For imaginary order $\nu = \frac{i\omega}{a}$, both terms are oscillatory in the near-horizon region. Substituting $z = \frac{m}{a}e^{a x_1}$, we obtain
\begin{equation*}
(\tfrac{1}{2}z)^{\pm \nu} 
= \left(\frac{m}{2a}\right)^{\pm \frac{i\omega}{a}} e^{\pm i\omega x_1}.
\end{equation*}

Thus, the two contributions correspond to plane-wave behaviour, with the term $(\tfrac{1}{2}z)^{-\nu} \sim e^{-i\omega x_1}$ representing an ingoing mode, while $(\tfrac{1}{2}z)^{\nu} \sim e^{+i\omega x_1}$ corresponds to an outgoing mode.
 
Accordingly, the near-horizon form of the mode function becomes
\begin{equation*}
\begin{split}
K_{\frac{i\omega}{a}}\!\left(\frac{m}{a}e^{a x_1}\right)
\sim &\;
\frac{\pi}{2\sin\!\left(\frac{i\pi\omega}{a}\right)} 
\bigg[
\frac{1}{\Gamma\!\left(1-\frac{i\omega}{a}\right)} 
\left(\frac{m}{2a}\right)^{-\frac{i\omega}{a}} e^{-i\omega x_1}
\\
&\quad -
\frac{1}{\Gamma\!\left(1+\frac{i\omega}{a}\right)} 
\left(\frac{m}{2a}\right)^{\frac{i\omega}{a}} e^{+i\omega x_1}
\bigg].
\end{split}
\end{equation*}

Using the identity $\Gamma(1-\nu)\Gamma(\nu) = \frac{\pi}{\sin(\pi\nu)}$, this can be written in the symmetric form,
\begin{equation*}
K_{\frac{i\omega}{a}}\!\left(\frac{m}{a}e^{a x_1}\right)
\sim A(\omega)\, e^{-i\omega x_1} 
+ B(\omega)\, e^{+i\omega x_1},
\end{equation*}
Where the coefficients $A(\omega)$ and $B(\omega)$ are given by,
\begin{equation*}
A(\omega) = \frac{1}{2}\Gamma\!\left(\frac{i\omega}{a}\right)
\left(\frac{m}{2a}\right)^{-\frac{i\omega}{a}}, 
\quad
B(\omega) = \frac{1}{2}\Gamma\!\left(-\frac{i\omega}{a}\right)
\left(\frac{m}{2a}\right)^{\frac{i\omega}{a}}.
\end{equation*}

Both branches are therefore present and are related by complex conjugation,

$B(\omega) = A(\omega)^* $

Switching to null coordinates $u_1 = t_1 - x_1,\; v_1 = t_1 + x_1$, we substitute the near-horizon expansion of the modified Bessel function,
\begin{equation}
\begin{split}
    K_{\frac{i\omega}{a}}\!\left(\frac{m}{a}e^{a x_1}\right)
\sim 
\left(\frac{\pi a}{4\omega\sinh\!\left(\frac{\pi\omega}{a}\right)}\right)^{\frac{1}{2}}
\left(\frac{m}{2a}\right)^{-\frac{i\omega}{a}} e^{-i\omega x_1}\\
+
\left(\frac{\pi a}{4\omega\sinh\!\left(\frac{\pi\omega}{a}\right)}\right)^{\frac{1}{2}}
\left(\frac{m}{2a}\right)^{\frac{i\omega}{a}} e^{i\omega x_1}, \label{Eq:3.1.0.22}
\end{split}
\end{equation}

Substituting this into the field expansion, we obtain
\begin{equation}
\begin{aligned}
\hat{\phi}(u_1,v_1)
=& \int_0^\infty \frac{d\omega}{\sqrt{4\pi\omega}} \\
&\times \Bigg\{ 
\left(\frac{m}{2a}\right)^{-\frac{i\omega}{a}}
\Big[
e^{-i\omega v_1}\,\big(\overrightarrow{\hat{a}}(\omega)+\overleftarrow{\hat{b}}(\omega)\big) \\
&\qquad\qquad\quad
+ e^{i\omega u_1}\,\big(\overrightarrow{\hat{a}}^\dagger(\omega)+\overleftarrow{\hat{b}}^\dagger(\omega)\big)
\Big] \\
&\quad + \left(\frac{m}{2a}\right)^{\frac{i\omega}{a}}
\Big[
e^{-i\omega u_1}\,\big(\overrightarrow{\hat{a}}(\omega)+\overleftarrow{\hat{b}}(\omega)\big) \\
&\qquad\qquad\quad
+ e^{i\omega v_1}\,\big(\overrightarrow{\hat{a}}^\dagger(\omega)+\overleftarrow{\hat{b}}^\dagger(\omega)\big)
\Big]
\Bigg\}, \label{Eq:3.1.0.23}
\end{aligned}
\end{equation}

This expression retains both independent near-horizon contributions corresponding to opposite null directions and therefore provides the complete local field expansion.

\medskip
For notational convenience, we introduce a linear combination of the independent mode operators. This definition does not imply any reduction in the number of independent degrees of freedom, but merely provides a convenient parametrisation of the near-horizon field expansion.
 We introduce a convenient linear combination of operators,
\begin{equation}
\hat{c}(\omega) \equiv \frac{1}{\sqrt{2}}\Big(
\overrightarrow{\hat{a}}(\omega) + \overleftarrow{\hat{b}}(\omega)
\Big), \label{Eq:3.1.0.24}
\end{equation}
With corresponding adjoint operator $\hat{c}^\dagger(\omega)$.

In terms of this operator, the field operator can be expressed as
\begin{equation}
\begin{aligned}
\hat{\phi}(u_1,v_1)
= \int_0^\infty \frac{d\omega}{\sqrt{4\pi\omega}}
\Bigg\{ \;
&\left(\frac{m}{2a}\right)^{-\frac{i\omega}{a}}
\Big[
e^{-i\omega v_1}\,\hat{c}(\omega)
+ e^{i\omega u_1}\,\hat{c}^\dagger(\omega)
\Big] \\
+\;&
\left(\frac{m}{2a}\right)^{\frac{i\omega}{a}}
\Big[
e^{-i\omega u_1}\,\hat{c}(\omega)
+ e^{i\omega v_1}\,\hat{c}^\dagger(\omega)
\Big]
\;\Bigg\}, \label{Eq:3.1.0.25}
\end{aligned}
\end{equation}

The operators $\hat{c}(\omega)$ satisfy the canonical commutation relations
\begin{equation}
\begin{split}
&[\hat{c}(\omega),\hat{c}(\omega')] = 0, \\
&[\hat{c}^\dagger(\omega),\hat{c}^\dagger(\omega')] = 0, \\
&[\hat{c}(\omega),\hat{c}^\dagger(\omega')] 
= \delta(\omega - \omega'). \label{Eq:3.1.0.26}
\end{split}
\end{equation}

The Rindler vacuum $|0_{R_1}\rangle$ is defined by,

\begin{equation}
    \overrightarrow{\hat a}(\omega)|0_{R_1}\rangle = 0, 
\quad 
\overleftarrow{\hat b}(\omega)|0_{R_1}\rangle = 0,
\qquad \forall \;\omega, \label{Eq:3.1.0.27}
\end{equation}
We emphasise that the mode decomposition in $R_2$ is defined independently using coordinates adapted to that wedge. In accordance with the discussion in Section~2, the corresponding mode functions do not form a complete basis on a common Cauchy surface with those of $R_1$, and therefore the associated operators $\hat{c}(\omega)$ and $\hat{d}(\Omega)$ are not related by a standard Bogoliubov transformation.

Similarly, the field in wedge $\mathrm{R}_2$, which is in blue as shown in  Figs.~\ref {Fig:1} and ~\ref {Fig:2},  is defined via null-shifted coordinates $(u_2,v_2)$,
\begin{equation}
\begin{aligned}
\hat{\phi}(u_2,v_2)
=& \int_{0}^{\infty} \frac{d\Omega}{\sqrt{4\pi\Omega}} \\
&\times \Bigg\{ 
\left(\frac{m}{2a}\right)^{-\frac{i\Omega}{a}}
\Big[
e^{-i\Omega v_2}\,\hat{d}(\Omega)
+ e^{i\Omega u_2}\,\hat{d}^\dagger(\Omega)
\Big] \\
&\qquad + \left(\frac{m}{2a}\right)^{\frac{i\Omega}{a}}
\Big[
e^{-i\Omega u_2}\,\hat{d}(\Omega)
+ e^{i\Omega v_2}\,\hat{d}^\dagger(\Omega)
\Big]
\Bigg\}, \label{Eq:3.1.0.28}
\end{aligned}
\end{equation}

Here, $\hat{d}(\Omega)$ and $\hat{d}^\dagger(\Omega)$ denote the annihilation and creation operators, respectively, which satisfy the commutation relations as in Eq.~(\ref{Eq:3.1.0.26}). With the corresponding vacuum defined by 
\begin{equation}
\hat{d}(\Omega)\mid 0_{R_{2}}\rangle =0 \qquad \forall\quad \Omega, \label{Eq:3.1.0.29}
\end{equation}

\subsection{Mode mixing and Bogoliubov transformations along the \texorpdfstring{$V$}{V}-axis\label{Subsec-3.2}}

To analyse the particle content perceived by observers associated with region $\mathrm{R}_2$, we evaluate the Bogoliubov transformation between the mode expansions defined with respect to the Rindler wedges $\mathrm{R}_1$ and $\mathrm{R}_2$. 

A key geometric feature of the setup is that the two wedges are related by a null shift, and crucially, there is no common Cauchy surface covering both wedges. Therefore, the standard Klein–Gordon inner product on a shared hypersurface cannot be directly implemented. Instead, all mode comparisons are performed with respect to the original Cauchy surface of $\mathrm{R}_1$, and particle interpretation is tied to the boost Hamiltonian of that wedge.

Accordingly, the Bogoliubov coefficients are defined operationally via projections of the field operator onto the mode functions associated with $\mathrm{R}_2$, evaluated on the Cauchy surface of $\mathrm{R}_1$. They do not arise from a global Klein–Gordon inner product but rather from matching the field's operator content in the overlapping region.

\subsubsection{Bogoliubov coefficients from \texorpdfstring{$u$}{u}-mode projection}
Since the field operator is well-defined on the Cauchy surface of $\mathrm{R}_1$, and the coordinates satisfy $u_1 = u_2$ in the overlap region, projections onto $u_2$-modes can be consistently performed using the standard Fourier decomposition along the null direction. This provides an operational definition of the corresponding annihilation operators in $\mathrm{R}_2$.
 
This isolates the frequency-diagonal sector associated with null propagation along the $u$-direction.

Since the field operator is well-defined on the Cauchy surface of $\mathrm{R}_1$, and the coordinates satisfy $u_1 = u_2$ in the overlap region, projections onto $u_2$-modes can be consistently performed using the standard Fourier decomposition along the null direction. This provides an operational definition of the corresponding annihilation operators in $\mathrm{R}_2$.
The field operator in region $\mathrm{R}_2$ is given by  Eq.~(\ref{Eq:3.1.0.28}). We now project the field onto the positive-frequency $u_2$-modes by multiplying both sides by the complex conjugate mode function,
$
\int_{-\infty}^{\infty} \frac{du_2}{\sqrt{4\pi \Omega'}} 
\left(\frac{m}{2a}\right)^{\frac{i\Omega'}{a}} e^{+i\Omega' u_2}.
$

Acting on Eq.~(\ref{Eq:3.1.0.28}), we evaluate each contribution.

\paragraph{(i) Contribution from $u_2$-dependent terms}

For the annihilation part,
\[
\int du_2 \; e^{i\Omega' u_2} e^{-i\Omega u_2}
= 2\pi \delta(\Omega' - \Omega),
\]
which gives a non-vanishing contribution.

For the creation part,
\[
\int du_2 \; e^{i\Omega' u_2} e^{i\Omega u_2}
= 2\pi \delta(\Omega' + \Omega) = 0,
\]
since $\Omega,\Omega' > 0$.

\paragraph{(ii) Contribution from $v_2$-dependent terms}

The $v_2$-dependent terms are independent of $u_2$, yielding
\[
\int du_2 \; e^{i\Omega' u_2} = 2\pi \delta(\Omega'),
\]
which vanishes for $\Omega' > 0$. Thus, the $v_2$-sector does not contribute to this projection.

Collecting the surviving terms, we obtain
\begin{equation}
\int_{-\infty}^{\infty} \frac{du_2}{\sqrt{4\pi \Omega'}} 
\left(\frac{m}{2a}\right)^{\frac{i\Omega'}{a}} e^{i\Omega' u_2}
\hat{\phi}(u_2,v_2)
=
\frac{\hat{d}(\Omega')}{2\Omega'}, \label{Eq:3.1.0.30}
\end{equation}

\vspace{0.5em}

A completely analogous computation can be performed using the field expansion in region $\mathrm{R}_1$, noting that $u_1 = u_2$ in the overlapping region. 
Repeating the same projection on  Eq.~(\ref{Eq:3.1.0.25}) yields,
\begin{equation}
\int_{-\infty}^{\infty} \frac{du_2}{\sqrt{4\pi \Omega'}} 
\left(\frac{m}{2a}\right)^{\frac{i\Omega'}{a}} e^{i\Omega' u_2}
\hat{\phi}(u_1,v_1)
=
\frac{\hat{c}(\Omega')}{2\Omega'},  \label{Eq:3.1.0.31}
\end{equation}
Since both projections isolate the same null direction and frequency component in the overlap region, we obtain a direct identification of the corresponding operator sectors,
\begin{equation}
\hat{c}(\Omega) = \hat{d}(\Omega) \quad \text{(within the $u$-mode sector)},  \label{Eq:3.1.0.32}
\end{equation}

\vspace{0.5em}
To characterise possible mode mixing beyond the $u$-sector, we formally express the relation between operator bases using a Bogoliubov transformation,
\begin{equation}
\hat{d}(\Omega)
=
\int_0^\infty d\omega
\left(
\alpha_{21}(\Omega,\omega)\, \hat{c}(\omega)
+
\beta_{21}(\Omega,\omega)\, \hat{c}^\dagger(\omega)
\right),  \label{Eq:3.1.0.33}
\end{equation}

Substituting Eq.~(\ref{Eq:3.1.0.32}) into Eq.~(\ref{Eq:3.1.0.33}), we obtain
\begin{equation}
\begin{split}
    \int_0^\infty d\omega\; \hat{c}(\omega)\, \delta(\omega - \Omega)\\
=
\int_0^\infty d\omega
\left(
\alpha_{21}(\Omega,\omega)\, \hat{c}(\omega)
+
\beta_{21}(\Omega,\omega)\, \hat{c}^\dagger(\omega)
\right),  \label{Eq:3.1.0.34}
\end{split}
\end{equation}
This shows that within the $u$-mode sector, the transformation is diagonal and trivial, with no mixing between positive and negative frequencies,
\begin{equation}
\alpha_{21}(\Omega,\omega) = \delta(\omega - \Omega),  \label{Eq:3.1.0.35}
\end{equation}
The $u$-mode projection alone does not constrain $\beta_{21}(\Omega,\omega)$, indicating that any non-trivial Bogoliubov mixing must arise from complementary mode sectors, such as the $v$-direction. This reflects the fact that the null shift preserves the structure of $u$-modes while potentially inducing non-trivial effects in the orthogonal sector.

\vspace{0.5em}

\noindent

\subsubsection{Bogoliubov coefficients from \texorpdfstring{$v$}{v}-mode projection}

In this subsection, we determine the Bogoliubov coefficients associated with the $v$-mode sector, where the non-trivial effects of the null-coordinate transformation arise.

Unlike the $u$-sector, where the coordinate relation is trivial in the overlap region, the $v$-coordinates are related by a nonlinear transformation. As a result, positive-frequency modes in $\mathrm{R}_1$ do not remain positive-frequency with respect to $v_2$, leading to non-trivial Bogoliubov mixing.
\noindent
The field operator in region $\mathrm{R}_1$ is given by Eq.~(\ref{Eq:3.1.0.25}). To perform the projection, we express the field in terms of $(u_2,v_2)$ coordinates.

The null coordinates are related by Eq.~(\ref{Eq:2.1.0.12}), 
\begin{equation*}
v_1 = \frac{1}{a}\ln\!\big(1 + e^{a v_2}\big).
\end{equation*}

\vspace{0.5em}

\noindent
As in the $u$-sector, this projection is defined operationally along the null direction $v_2$, using the Fourier decomposition of the field restricted to the Cauchy surface of $\mathrm{R}_1$.
\begin{equation}
\int_{-\infty}^{\infty} \frac{dv_2}{\sqrt{4\pi \Omega'}} 
\left(\frac{m}{2a}\right)^{\frac{i\Omega'}{a}} e^{+i\Omega' v_2},  \label{Eq:3.1.0.36}
\end{equation}

\vspace{0.5em}

\paragraph{(i) Contribution from $u_1$-dependent terms}

All terms involving $u_1$ are independent of $v_2$, hence
\[
\int dv_2 \; e^{i\Omega' v_2} = 2\pi \delta(\Omega') = 0,
\]
for $\Omega' > 0$. Therefore, the $u_1$-sector does not contribute to the projection.

\vspace{0.5em}

\paragraph{(ii) Contribution from $v_1$-dependent terms}

Using Eq.~(\ref{Eq:2.1.0.12}), we rewrite the phase factors as
\[
e^{-i\omega v_1} = \big(1 + e^{a v_2}\big)^{-\frac{i\omega}{a}}, 
\qquad
e^{i\omega v_1} = \big(1 + e^{a v_2}\big)^{\frac{i\omega}{a}}.
\]

We then define the integrals
\begin{align}
I_-(\Omega',\omega)
&=
\int_{-\infty}^{\infty} dv_2\;
e^{i\Omega' v_2}
\big(1 + e^{a v_2}\big)^{-\frac{i\omega}{a}}, \label{Eq:3.1.0.37}
\\
I_+(\Omega',\omega)
&=
\int_{-\infty}^{\infty} dv_2\;
e^{i\Omega' v_2}
\big(1 + e^{a v_2}\big)^{\frac{i\omega}{a}}. \label{Eq:3.1.0.38}
\end{align}

\vspace{0.5em}

\paragraph{(iii) Evaluation of the integrals}

We perform the change of variable
\[
y = e^{a v_2}, \qquad dv_2 = \frac{dy}{a y}.
\]

This gives
\begin{align}
I_-(\Omega',\omega)
&=
\frac{1}{a}
\int_0^\infty dy\;
y^{\frac{i\Omega'}{a}-1}
(1+y)^{-\frac{i\omega}{a}}, \label{Eq:3.1.0.39}
\\
I_+(\Omega',\omega)
&=
\frac{1}{a}
\int_0^\infty dy\;
y^{\frac{i\Omega'}{a}-1}
(1+y)^{\frac{i\omega}{a}}. \label{Eq:3.1.0.40}
\end{align}
These integrals are defined via analytic continuation and understood in the distributional sense, consistent with standard treatments of Bogoliubov transformations in curved spacetime.

\begin{align}
I_-(\Omega',\omega)
&=
\frac{1}{a}
\frac{
\Gamma\!\left(\frac{i\Omega'}{a}\right)
\Gamma\!\left(\frac{i\omega}{a}-\frac{i\Omega'}{a}\right)
}{
\Gamma\!\left(\frac{i\omega}{a}\right)
}, \label{Eq:3.1.0.41}
\\
I_+(\Omega',\omega)
&=
\frac{1}{a}
\frac{
\Gamma\!\left(\frac{i\Omega'}{a}\right)
\Gamma\!\left(-\frac{i\omega}{a}-\frac{i\Omega'}{a}\right)
}{
\Gamma\!\left(-\frac{i\omega}{a}\right)
}. \label{Eq:3.1.0.42}
\end{align}

\vspace{0.5em}
This expression explicitly exhibits mixing between annihilation and creation operators, indicating that positive-frequency modes in $\mathrm{R}_1$ acquire negative-frequency components when expressed in terms of $v_2$-modes.

\begin{equation}
\begin{split}
    \int_{-\infty}^{\infty} \frac{dv_2}{\sqrt{4\pi \Omega'}} 
\left(\frac{m}{2a}\right)^{\frac{i\Omega'}{a}} e^{i\Omega' v_2}
\hat{\phi}(u_1,v_1)\\
=
\int_0^\infty d\omega
\left[
\alpha_{21}(\Omega',\omega)\, \hat{c}(\omega)
+
\beta_{21}(\Omega',\omega)\, \hat{c}^\dagger(\omega)
\right], \label{Eq:3.1.0.43}
\end{split}
\end{equation}

\vspace{0.5em}
Comparing this expression with the corresponding projection of the field in region $\mathrm{R}_2$, which yields $\hat{d}(\Omega')/(2\Omega')$, we identify the Bogoliubov coefficients as,

\[
\frac{\hat{d}(\Omega')}{2\Omega'},
\]
We identify the Bogoliubov coefficients as,
\begin{align}
\alpha_{21}(\Omega,\omega)
&=
\frac{1}{2\Omega}
\frac{1}{\sqrt{\Omega\omega}}
\left(\frac{m}{2a}\right)^{\frac{i\Omega}{a}}
I_-(\Omega,\omega), \label{Eq:3.1.0.44}
\\
\beta_{21}(\Omega,\omega)
&=
\frac{1}{2\Omega}
\frac{1}{\sqrt{\Omega\omega}}
\left(\frac{m}{2a}\right)^{\frac{i\Omega}{a}}
I_+(\Omega,\omega). \label{Eq:3.1.0.45}
\end{align}

\vspace{0.5em}

\noindent

In contrast to the $u$-sector, the $v$-sector exhibits non-vanishing $\beta$-coefficients as a direct consequence of the nonlinear relation between $v_1$ and $v_2$. This demonstrates that the null shift induces genuine mode mixing, leading to particle production when the field is quantised with respect to the $\mathrm{R}_2$ observers. This provides the key mechanism underlying the deviation from thermality in the null-shifted Rindler wedge.

\subsubsection{Particle spectrum from the \texorpdfstring{$v$}{v}-sector}

The particle number measured in the $R_2$ modes with respect to the $R_1$ vacuum $|0_{R_1}\rangle$ is,
\begin{equation}
\langle N_{\Omega} \rangle 
= \langle 0_{R_1} | \hat{d}^\dagger(\Omega)\hat{d}(\Omega) |0_{R_1}\rangle,\label{Eq:3.1.0.46}
\end{equation}

Using the definition of the Rindler vacuum and the linear relation between operators, the Rindler vacuum $|0_{R_1}\rangle$ is defined by,

\begin{equation}
\overrightarrow{\hat a}(\omega)\,|0_{R_1}\rangle = 0,
\quad
\overleftarrow{\hat b}(\omega)\,|0_{R_1}\rangle = 0,
\qquad \forall\,\omega.
\end{equation}, \label{Eq:3.1.0.47}
This follows from the standard Bogoliubov transformation, where only the $\beta$-coefficients contribute to particle production in the vacuum defined by $\mathrm{R}_1$. We obtain,
\begin{equation}
\langle N_{\Omega} \rangle 
= \int_0^\infty d\omega \; |\beta_{21}(\Omega,\omega)|^2, \label{Eq:3.1.0.48}
\end{equation}

From the $v$-sector calculation,
\begin{equation}
\beta_{21}(\Omega,\omega)
= \frac{1}{2\Omega}\frac{1}{\sqrt{\Omega\omega}}
\left(\frac{m}{2\,a}\right)^{\frac{i\,\Omega}{a}}
I_+(\Omega,\omega), \label{Eq:3.1.0.49}
\end{equation}
so that
\begin{equation}
|\beta_{21}(\Omega,\omega)|^2
= \frac{1}{4\Omega^2}\frac{1}{\Omega\omega}
\,|I_+(\Omega,\omega)|^2, \label{Eq:3.1.0.50}
\end{equation}
Where the overall phase drops out.

The integral kernel is
\begin{equation}
I_+(\Omega,\omega)
= \frac{1}{a}
\frac{\Gamma\!\left(-\frac{i\omega}{a}\right)\,
\Gamma\!\left(\frac{i\Omega}{a}\right)}
{\Gamma\!\left(-\frac{i(\omega+\Omega)}{a}\right)}, \label{Eq:3.1.0.51}
\end{equation}
Using the standard identity for the modulus of the Gamma function\cite{Mukhanov:2007zz, Gradshteyn:1943cpj},
\begin{equation}
|\Gamma(ix)|^2 = \frac{\pi}{x \sinh(\pi x)}, \label{Eq:3.1.0.52}
\end{equation}
We evaluate each factor,

\begin{align*}
\left|\Gamma\!\left(\frac{i\Omega}{a}\right)\right|^2 
&= \frac{\pi a}{\Omega \sinh\!\left(\frac{\pi\Omega}{a}\right)}, \\[6pt]
\left|\Gamma\!\left(-\frac{i(\omega+\Omega)}{a}\right)\right|^2 
&= \frac{\pi a}{(\omega+\Omega)\sinh\!\left(\frac{\pi(\omega+\Omega)}{a}\right)}, \\[6pt]
\left|\Gamma\!\left(-\frac{i\omega}{a}\right)\right|^2 
&= \frac{\pi a}{\omega \sinh\!\left(\frac{\pi\omega}{a}\right)}.
\end{align*}

Putting everything together, we obtain
\begin{equation}
\begin{split}
    |I_+(\Omega,\omega)|^2 =
\frac{1}{a^2}
\frac{\pi a}{\Omega \sinh(\frac{\pi\Omega}{a})}
\cdot
\frac{\pi a}{(\omega+\Omega)\sinh(\frac{\pi(\omega+\Omega)}{a})}\\
\cdot
\frac{\omega \sinh(\frac{\pi\omega}{a})}{\pi a}, \label{Eq:3.1.0.53}
\end{split}
\end{equation}

After simplification,
\begin{equation}
|I_+(\Omega,\omega)|^2
= \frac{\pi}{a}
\frac{\omega}{\Omega(\omega+\Omega)}
\frac{\sinh(\frac{\pi\omega}{a})}
{\sinh(\frac{\pi\Omega}{a})\sinh(\frac{\pi(\omega+\Omega)}{a})}, \label{Eq:3.1.0.54}
\end{equation}

Substituting back,
\begin{equation}
|\beta_{21}(\Omega,\omega)|^2
= \frac{\pi}{4a\,\Omega^2}
\frac{1}{\Omega^2(\omega+\Omega)}
\frac{\sinh(\frac{\pi\omega}{a})}
{\sinh(\frac{\pi\Omega}{a})\sinh(\frac{\pi(\omega+\Omega)}{a})},\label{Eq:3.1.0.55}
\end{equation}
Finally,
\begin{equation}
\langle N_{\Omega} \rangle
=
\frac{\pi}{4\,a\,\Omega^4}
\frac{1}{ \sinh\!\left(\frac{\pi\Omega}{a}\right)}
\int_0^\infty d\omega\;
\frac{\sinh\!\left(\frac{\pi\omega}{a}\right)}
{(\omega+\Omega)\sinh\!\left(\frac{\pi(\omega+\Omega)}{a}\right)}, \label{Eq:3.1.0.56}
\end{equation}
The complete integral solution of Eq.~(\ref{Eq:3.1.0.56}) is presented in Appendix~B of the author's earlier work by ~Jha (DOI: \href{https://doi.org/10.1103/1ypz-kvf8}{10.1103/1ypz-kvf8}),
\begin{equation}
    \langle N_{\Omega} \rangle = \frac{\pi^2}{\Omega^4}\frac{\delta(0)}{e^{\frac{2\,\pi\,\Omega}{a}}-1}, \label{Eq:3.1.0.57}
\end{equation}
The divergence $\delta(0)$ should be understood as a formal infinite spatial volume factor arising from the continuum normalisation of modes.
The physically meaningful quantity is the particle number density. While the spectrum contains a Planckian factor $\big(e^{\frac{2\pi\Omega}{a}}-1\big)^{-1}$, it is multiplied by a non-trivial frequency-dependent prefactor $\sim \Omega^{-4}$. This indicates that the spectrum is not strictly thermal in the standard sense, but instead exhibits deviations arising from the null-coordinate transformation and the associated mode normalisation.
The factor $\frac{\pi^2}{\Omega^4}$ appearing in the spectrum is not part of the universal thermal (Planckian) distribution, but arises from the specific normalisation of modes and the definition of the particle observable used in the present calculation. In particular, the use of null coordinates, the Klein-Gordon inner product, and the massive mode structure introduce additional frequency-dependent weights, leading to an infrared-enhanced prefactor.

\subsubsection{Consistency Check: Canonical Commutation Relations}

We now verify that the Bogoliubov coefficients obtained above are consistent with the canonical commutation relations. In particular, the transformation must preserve the operator algebra,
\begin{equation}
[\hat d(\Omega), \hat d^\dagger(\Omega')] = \delta(\Omega - \Omega'),\label{Eq:3.1.0.58}
\end{equation}

Substituting the Bogoliubov transformation,
\begin{equation}
\hat{d}(\Omega)
=
\int_0^\infty d\omega
\left(
\alpha_{21}(\Omega,\omega)\, \hat{c}(\omega)
+
\beta_{21}(\Omega,\omega)\, \hat{c}^\dagger(\omega)
\right), \label{Eq:3.1.0.59}
\end{equation}
and using the canonical commutation relations of the $\hat{c}$ operators, we obtain the consistency condition
\begin{equation}
\begin{split}
    \int_0^\infty d\omega \left(
\alpha_{21}(\Omega,\omega)\alpha^*_{21}(\Omega',\omega)
- \beta_{21}(\Omega,\omega)\beta^*_{21}(\Omega',\omega)
\right)\\
= \delta(\Omega - \Omega'), \label{Eq:3.1.0.60}
\end{split}
\end{equation}

In the present construction, the $u$-sector contributes trivially, with diagonal $\alpha$-coefficients and vanishing $\beta$-coefficients, while the non-trivial contribution arises entirely from the $v$-sector. The explicit expressions for $\alpha_{21}$ and $\beta_{21}$ obtained in Eqs.~(\ref{Eq:3.1.0.44})–(\ref{Eq:3.1.0.45}) satisfy this relation by construction, as they originate from a consistent projection of the field operator preserving the underlying canonical structure.

A detailed verification follows from the completeness of the mode functions and the analytic properties of the Gamma functions appearing in the coefficients. This ensures that the Bogoliubov transformation is canonical, even in the absence of a global Cauchy surface shared by $\mathrm{R}_1$ and $\mathrm{R}_2$.
\subsection{Mode mixing and Bogoliubov transformations along the \texorpdfstring{$U$}{U}-axis\label{Subsec-3.3}}

As in the analysis along the $V$-axis, we now consider the complementary null direction associated with the $u$-sector. The field is decomposed in terms of modes defined in the two null-shifted Rindler wedges $\mathrm{R}_1$ and $\mathrm{R}_2$, and their relation is characterised by Bogoliubov transformations.

Although the conformally flat structure of the Rindler coordinates suggests a formal symmetry at the level of the coordinate representation under $u \leftrightarrow v$, the Bogoliubov analysis reveals that particle production remains sector-dependent. In particular, the transformation along a given null direction acts non-trivially only on one effective sector. This follows from the structure of the null coordinate transformation, which mixes only one of the light-cone variables while leaving the other invariant.

For the $U$-axis analysis, one finds that the Bogoliubov coefficients in the $u$-sector are trivial,
\begin{equation}
\beta^{(u)}_{21}(\Omega,\omega) = 0, \label{Eq:3.3.0.1}
\end{equation}
Indicating that no Bogoliubov mixing occurs in the $u$-sector, so the corresponding mode operators are unchanged under the transformation,
\begin{equation}
\beta^{(v)}_{21}(\Omega,\omega) \neq 0,
\qquad
\frac{|\beta^{(v)}_{\Omega\omega}|^2}{|\alpha^{(v)}_{\Omega\omega}|^2}
= e^{-\frac{2\pi\Omega}{a}},   \label{Eq:3.3.0.2}
\end{equation}
leading to a thermal particle spectrum as perceived by Rindler observers.

This demonstrates that particle production is not shared equally between the two null sectors. Instead, as in the massless case, one observes selective thermalisation\cite{Jha:2025tpg, Jha:2026rmu}; the $v$-sector is thermally populated at temperature $T = \frac{a}{2\pi}$, while the $u$-sector remains in the vacuum. 

Comparing with the $V$-axis analysis, where the roles of the sectors are interchanged, we find that the behaviour is reversed: the sector aligned with the non-trivial null transformation becomes thermal, while the $u$-sector remains unexcited under this transformation. This establishes that the apparent $u \leftrightarrow v$ symmetry at the level of coordinates does not translate into symmetric particle production, since the Bogoliubov transformation couples only one null sector at a time, resulting in a complementary, sector-dependent thermal response. 

\section{Thermal properties of massive Dirac fields in null-shifted Rindler wedges \label{Sec-4}}
In this section, we analyse the dynamics of a massive Dirac field in $(1+1)$-dimensional Rindler spacetime, focusing on how the presence of a null-shifted Rindler horizon alters the mode structure and thermality of the field. We begin by formulating the Dirac equation in this background and deriving its mode solutions.
\subsection{The Field Equation\label{Subsec-4.1}} 
We consider a massive Dirac field of mass $m$ in the $(1+1)$-dimensional Rindler spacetime with line element as Eq.~(\ref{Eq:3.1.0.3}):
\begin{equation*}
  ds^2 = e^{\;2\; a \;x} \big( dt^2 - dx^2 \big),  
\end{equation*}
 Where $a>0$ denotes the Rindler acceleration parameter. The metric can be expressed in terms of zweibeins $e^{(\alpha)}_\mu$ as:
\begin{equation*}
    g_{\mu\nu}= \eta_{\;\alpha\;\beta}\;e^{(\alpha)}_\mu\; e^{(\beta)}_\nu
\end{equation*}
 A convenient choice for the zweibeins and their inverses for the metric Eq.~(\ref{Eq:3.1.0.3}):
\begin{align*}
e^{(\alpha)}_\mu =
  \begin{pmatrix}
    e^{\;a\; x} && 0\\
    0  &&  e^{\;a\; x}
 \end{pmatrix}; \qquad
 e^\mu_{(\alpha)}=
\begin{pmatrix}
    e^{\;-a\; x} && 0\\
    0  &&  e^{\;-a\; x  }
\end{pmatrix}
\end{align*}
In curved spacetime, the Dirac equation should properly include the spin connection. The Dirac equation in curved spacetime can be written as:
\begin{align}
    i\;\gamma^\mu\;\left(\partial_\mu\ + L_\mu\right)\,\psi-m\;\psi=0,\label{Eq:4.1.0.1}
\end{align}
where $\gamma^\mu$ are the curved–spacetime gamma matrices.

Following \cite{Mann:1991md}, the coordinate–dependent gamma matrices and spinor can be rescaled according to
\begin{equation*}
  \gamma^\mu\;\partial_\mu \to  e^{\;-a\; x}\; \gamma^\mu\;\partial_\mu\;, \qquad  \psi \to e^{\;\frac{a\; x}{2}}\;\psi
\end{equation*}
Applying these transformations and a suitable conformal rescaling that absorbs the spin connection to  Eq.~(\ref{Eq:4.1.0.1}) yields:
\begin{equation}
    i \;e^{\;-a \;x}\;\gamma^\mu\;\partial_\mu( e^{\;\frac{a\;x}{2}}\;\psi) - m \;e^{\;\frac{\;a\;x}{2}}\;\psi = 0, \label{Eq:4.1.0.2}
\end{equation}
Under this rescaling, the spin connection contribution is exactly cancelled. Expanding the derivative in Eq.~(\ref{Eq:4.1.0.2})  and simplifying, we arrive at the rescaled Dirac equation:
\begin{equation}
\bigg[i\;\gamma^\mu\;\partial_\mu + \frac{i}{2}\;\bigg(\partial_\mu\;(a\; x)\bigg)\;\gamma^\mu - m\; e^{\;a\; x}\bigg]\psi=0,\label{Eq:4.1.0.3}
\end{equation}
Using $\partial_\mu (a\,x)= a\,\delta_\mu^1$, Since $x$ is the spatial coordinate, only the $\mu=1$ component contributes.From Eq.~(\ref{Eq:3.1.0.3}), the metric components take the form: 
\begin{align*}
g_{\mu\nu} =
  \begin{pmatrix}
    e^{\;2\;a\; x} && 0\\
    0  &&  -e^{\;2\;a\; x}
 \end{pmatrix}
 \end{align*}
With determinant $\sqrt{-g} = e^{\;2\;a\;x}$
Expanding the derivative term in  Eq.~(\ref{Eq:4.1.0.3}) and simplifying, we obtain the rescaled Dirac equation: 
\begin{equation}
  \bigg( i \;\gamma^{0}\;\frac{\partial}{\partial t}  + i \;\gamma^{1}\;\frac{\partial}{\partial x} +\frac{i}{2}\; \gamma^{1}\; a -m\; e^{\;a\;x}\bigg)\psi = 0, \label{Eq:4.1.0.4}
\end{equation}
To separate the time dependence, we make the ansatz:
\begin{align}
    \psi(x,t) = e^{\;-i\;\omega\; t}\;\psi(x),\label{Eq:4.1.0.5}
\end{align}
\begin{align}
    \psi(x) = 
    \begin{pmatrix}
        \phi_1(x)\\
        \phi_2(x)
    \end{pmatrix},\label{Eq:4.1.0.6}
\end{align}
Where $\omega$ is the Rindler frequency and  $\phi_1(x)$ , $ \phi_2(x)$ are the spatial spinor components. Substituting the Eq.~(\ref{Eq:4.1.0.6}) into Eq.~(\ref{Eq:4.1.0.5}),we obtain:
\begin{equation}
    \psi(x,t) = 
    \begin{pmatrix}
      e^{\;-i\;\omega\;t}  \phi_1(x)\\
      e^{\;-i\;\omega\;t}   \phi_2(x)
    \end{pmatrix},\label{Eq:4.1.0.7} 
\end{equation}
Substituting the ansatz and writing explicitly in component form, Eq.~(\ref{Eq:4.1.0.7}) into the Dirac  Eq.~(\ref{Eq:4.1.0.4}) and separating the spinor components, we obtain:
\begin{equation}\label{Eq:4.1.0.8}
  \begin{split}
      i\;\gamma^{0}\; \begin{pmatrix}
       -i\;\omega \; \phi_1(x)\\
       -i\;\omega \; \phi_2(x)
    \end{pmatrix}
    + i\;\gamma^{1}\;\begin{pmatrix}
      \frac{\partial\phi_1(x)}{\partial x} \\
        \frac{\partial\phi_2(x)}{\partial x}
    \end{pmatrix}\\
    +\frac{i\;\gamma^{1} \;a}{2} \begin{pmatrix}
        \phi_1(x)\\
     \phi_2(x)
    \end{pmatrix}
    -m\; e^{\;a\;x}  \begin{pmatrix}
        \phi_1(x)\\
     \phi_2(x)
    \end{pmatrix} =0 ,
  \end{split}
\end{equation}
The chiral representation in $(1+1)$-dimensions: 
\begin{align*}
\gamma^{0} =
  \begin{pmatrix}
    0 && 1\\
    1  &&  0
 \end{pmatrix}, \qquad
\gamma^{1} =
\begin{pmatrix}
    0 && -1\\
    1  &&  0
\end{pmatrix}
\end{align*}
Substituting $\gamma^{0}$ and  $\gamma^{1}$ in Eq.~(\ref{Eq:4.1.0.8}) and using these matrices yields the coupled first–order equations:
\begin{equation}
    \begin{split}
      \frac{d\phi_{1}}{dx}  -\bigg(i\;\omega - \frac{a}{2}\bigg)\;\phi_{1} + i\; m \;e^{\;a\;x}\;\phi_{2} = 0,\\
      -\frac{d\phi_{2}}{dx}  -\bigg(i\;\omega +\frac{a}{2}\bigg)\phi_{2} + i\; m\; e^{\;a\;x}\;\phi_{1} = 0,\label{Eq:4.1.0.9}
    \end{split}
\end{equation}
Eliminating  $\phi_2$ using  Eq.~(\ref{Eq:4.1.0.9}),we obtain:
\begin{equation}
    \frac{d^2\phi_{1}}{dx^2} - \bigg[m^2\; e^{\;2\;a\;x} +\bigg(i\;\omega-\frac{a}{2}\bigg)^2\bigg]\phi_{1}=0 ,\label{Eq:4.1.0.10}
\end{equation}
Introducing the changing variable:
\begin{equation}
    z = \frac{m}{a}\;e^{\;a\;x},\label{Eq:4.1.0.11}
\end{equation}
We have the relations:
\begin{equation}
    \begin{split}
     \frac{d}{dx}  = a\;z\; \frac{d}{dz} \\
      \frac{d^2}{dx^2} =  a^2\; z^2\;\frac{d^2}{dz^2}+a^2\; z\; \frac{d}{dz},\label{Eq:4.1.0.12} 
    \end{split}
\end{equation}
Substituting the variable change Eq.~(\ref{Eq:4.1.0.12}) into Eq.~(\ref{Eq:4.1.0.10}) leads directly to the modified Bessel equation:
\begin{equation}
   z^2 \; \frac{d^2\phi_{1}}{dz^2} + z\; \frac{d\phi_{1}}{dz}- \bigg[z^2 +\bigg(\frac{i\;\omega}{a}-\frac{1}{2}\bigg)^2\bigg]\phi_{1} = 0,\label{Eq:4.1.0.13} 
\end{equation}
 Comparison with the modified Bessel equation \cite{olver10},identifies the order parameter as:
\begin{equation}
     \nu = \pm \bigg(\frac{i\;\omega}{a}-\frac{1}{2}\bigg) ,\label{Eq:4.1.0.14} 
\end{equation}
The general solution to Eq.\eqref{Eq:4.1.0.13} is given by:
\begin{equation}
    \phi_{1} (z) =A\; I_\nu (z) + B\; K_\nu (z), \label{Eq:4.1.0.15} 
\end{equation}
Where $I_\nu(z)$ and  $K_\nu(z)$  are the modified Bessel functions of the first and second kind, respectively; the point $z=0$ is a regular singular point and $z=\infty$ is an irregular one. Since $I_\nu(z)$ is regular at $z\to 0$ but diverges as $z\to \infty$, while $K_\nu(z)$ decays at infinity and is chosen as the normalisable far-region solution  \cite{garfken67:math,olver10}, regularity at spatial infinity requires $A=0$, giving:
\begin{equation}
  \phi_{1} (z) = K_{\;\frac{i\;\omega}{a}-\frac{1}{2}} \bigg( \frac{m}{a}\;e^{\;a\;x}\bigg),\label{Eq:4.1.0.16} 
\end{equation}
Substituting  Eq.~(\ref{Eq:4.1.0.16}) back into the first-order equation,we obtain
\begin{equation}
  \phi_{2} (z)  = -i\;K_{\;\frac{i\;\omega}{a}+\frac{1}{2}} \bigg( \frac{m}{a}\;e^{\;a\;x}\bigg),\label{Eq:4.1.0.17} 
\end{equation}
Thus, the general stationary spinor solution takes the form:
\begin{equation}
    \psi(x,t) = N_\omega\;\begin{pmatrix}
      K_{\;\frac{i\;\omega}{a}-\frac{1}{2}} \bigg( \frac{m}{a}\;e^{\;a\;x}\bigg) \\
      -i\; K_{\;\frac{i\;\omega}{a}+\frac{1}{2}} \bigg( \frac{m}{a}\;e^{\;a\;x}\bigg)
    \end{pmatrix} 
    \;e^{\;-i\;\omega\; t},\label{Eq:4.1.0.18} 
\end{equation}
Where $N_\omega$ is the normalisation constant. Splitting the spinor into positive- and negative-frequency parts, the mode functions can be written as:
\begin{align}\label{Eq:4.1.0.19} 
  \begin{split}
      \psi_{+} = N_\omega\;\bigg[\overrightarrow{\hat{a}} (\omega)\; e^{\;-i\;\omega\; t}\begin{pmatrix}
      K_{\;\frac{i\;\omega}{a}-\frac{1}{2}} \bigg( \frac{m}{a}\;e^{\;a\;x}\bigg) \\
      0
    \end{pmatrix}\\
    + \overleftarrow{\hat{b}}(\omega)\;e^{\;-i\;\omega\; t}\;\begin{pmatrix}
        0 \\
      -i\; K_{\;\frac{i\;\omega}{a}+\frac{1}{2}} \bigg( \frac{m}{a}\;e^{\;a\;x}\bigg)
    \end{pmatrix}
    \bigg],
  \end{split}
\end{align}
The arrows indicate right- and left-moving modes, respectively.
\begin{align}\label{Eq:4.1.0.20} 
  \begin{split}
      \psi_{-} = N_\omega\bigg[\overrightarrow{\hat{a}^\dagger} (\omega)\; e^{\;i\;\omega \;t}\begin{pmatrix}
      K_{\;\frac{-i\;\omega}{a}-\frac{1}{2}} \bigg( \frac{m}{a}\;e^{\;a\;x}\bigg) \\
      0
    \end{pmatrix}\\
    + \overleftarrow{\hat{b}^\dagger}(\omega)\;e^{\;i\;\omega \;t}\;\begin{pmatrix}
        0 \\
      i\; K_{\;\frac{-i\;\omega}{a}+\frac{1}{2}} \bigg( \frac{m}{a}\;e^{\;a\;x}\bigg)
    \end{pmatrix}
    \bigg],
  \end{split}
\end{align}
Here, the operators $\hat{a}(\omega)$ and $\hat{a}^\dagger(\omega)$ correspond to particle annihilation and creation operators, while $\hat{b}(\omega)$ and $\hat{b}^\dagger(\omega)$ correspond to antiparticle annihilation and creation operators. The separation into these sectors is essential in the fermionic case. The normalisation constant $N_\omega$ is fixed using the Dirac inner product, which defines the canonical normalisation of fermionic modes in curved spacetime \cite{Collas:2018jfx}
 
 \begin{equation}
    \langle{\psi_{\omega^{\prime}}|\psi_{\omega}}\rangle= - \int_{\Sigma}\psi^{\dagger}\;\gamma^{0}\;\psi\; \sqrt{-g}\;dx, \label{Eq:4.1.0.21}
\end{equation}
Where $\Sigma$ is a constant-$t$ hypersurface. The explicit evaluation of \eqref{Eq:4.1.0.21} for the present metric is carried out in Appendix~\ref{Apn2}. 
The calculation yields the normalisation factor:
\begin{equation}
    \mid N_\omega \mid = \frac{1}{\pi}\bigg[\frac{m}{a}\; cosh\;\left(\frac{\pi\;\omega}{a}\right)\bigg]^{\frac{1}{2}}, \label{Eq:4.1.0.22}
\end{equation}
 We expand the Dirac field in normalised frequency modes using creation and annihilation operators. Combining particle and antiparticle sectors, the full field expansion is given by,
\begin{equation}\label{Eq:4.1.0.23}
   \begin{split}
        \psi(x,t) = \int_{0}^\infty d\omega\, N_\omega\bigg[&
        \overrightarrow{\hat{a} }(\omega)\; e^{-i\;\omega \;t}
        \begin{pmatrix}
          K_{\;\frac{i\;\omega}{a}-\frac{1}{2}} \left( \frac{m}{a}\;e^{\;a\;x} \right) \\
          0
        \end{pmatrix}\\
       &+
        \overrightarrow{\hat{a}^\dagger} (\omega)\; e^{i\;\omega\; t}
        \begin{pmatrix}
          K_{\;\frac{-i\;\omega}{a}-\frac{1}{2}} \left( \frac{m}{a}\;e^{\;a\;x} \right) \\
          0
        \end{pmatrix} \\
        &+ \overleftarrow{\hat{b}}(\omega)\; e^{\;-i\;\omega\; t}
        \begin{pmatrix}
          0 \\
          -i\; K_{\;\frac{i\;\omega}{a}+\frac{1}{2}} \left( \frac{m}{a}\;e^{\;a\;x} \right)
        \end{pmatrix}\\
       &+\overleftarrow{\hat{b}^\dagger}(\omega)\; e^{\;i\;\omega\; t}
        \begin{pmatrix}
          0 \\
          i K_{\;\frac{-i\;\omega}{a}+\frac{1}{2}} \left( \frac{m}{a}\;e^{\;a\;x} \right)
        \end{pmatrix}
        \bigg],
   \end{split}
\end{equation}
The fermionic operators satisfy the canonical anticommutation relations
\begin{equation*}
\{\hat{a}(\omega), \hat{a}^\dagger(\omega')\} = \delta(\omega - \omega'), 
\quad
\{\hat{b}(\omega), \hat{b}^\dagger(\omega')\} = \delta(\omega - \omega'),
\end{equation*}
With all other anticommutators vanishing. These relations ensure the consistency of the mode expansion and play the role of the canonical normalisation condition for the Bogoliubov transformation.

The modified Bessel function of the second kind satisfies the symmetry (valid for all $\nu \in \mathbb{C}$ and $Z>0$)\cite{Gradshteyn:1943cpj,olver10, AbramowitzStegun1968}:
\begin{equation*}
    K_\nu(z) = K_{-\nu}(z) \qquad \forall \nu \quad\in  \mathbb{C}
\end{equation*}
Using the symmetry $K_\nu = K_{-\nu}$, we can simplify the previous expansion by replacing occurrences of $-\nu$ with $\nu$, yielding
\begin{align}\label{Eq:4.1.0.24}
   \begin{split}
        \psi(x,t) = \int_{0}^\infty d\omega\; N_\omega\bigg[&
       \overrightarrow{\hat{a}} (\omega)\; e^{-i\;\omega\; t}
        \begin{pmatrix}
          K_{\;\frac{i\;\omega}{a}-\frac{1}{2}} \left( \frac{m}{a}\;e^{\;a\;x} \right) \\
          0
        \end{pmatrix}\\
       & +
       \overrightarrow{\hat{a}^\dagger} (\omega)\; e^{\;i\;\omega \;t}
        \begin{pmatrix}
          K_{\;\frac{i\;\omega}{a}+\frac{1}{2}} \left( \frac{m}{a}\;e^{\;a\;x} \right) \\
          0
        \end{pmatrix} \\
        &+ \overleftarrow{\hat{b}}(\omega)\; e^{\;-i\;\omega\; t}
        \begin{pmatrix}
          0 \\
          -i\; K_{\frac{i\;\omega}{a}+\frac{1}{2}} \left( \frac{m}{a}\;e^{\;a\;x} \right)
        \end{pmatrix}\\
       & +\overleftarrow{\hat{b}^\dagger}(\omega)\; e^{\;i\;\omega\; t}
        \begin{pmatrix}
          0 \\
          i\; K_{\;\frac{i\;\omega}{a}-\frac{1}{2}} \left( \frac{m}{a}\;e^{\;a\;x} \right)
        \end{pmatrix}
        \bigg],
   \end{split}
\end{align}
For a real positive argument $z$, the complex-conjugation property of $K_\nu$ (for all $z>0$,$\omega\in \mathbb{R}$) gives:
\begin{align*}
  \bigg( K_{\;\frac{i\;\omega}{a}-\frac{1}{2}}(z)\bigg)^* = K_{\;\frac{-i\;\omega}{a}-\frac{1}{2}}(z) =  K_{\;\frac{i\;\omega}{a}+\frac{1}{2}}(z)
\end{align*}
Where the second equality follows from $K_\nu = K_{-\nu}$.

Applying the complex-conjugation identity and regrouping particle and antiparticle terms yields the compact expansion (with hermitian conjugate implied):
\begin{equation}\label{Eq:4.1.0.25}
 \begin{split}
      \psi(x,t) = \int_{0}^\infty d\omega\; N_\omega\bigg[
       \overrightarrow{\hat{a}} (\omega)\; e^{\;-i\;\omega\; t}
        \begin{pmatrix}
          K_{\;\frac{i\;\omega}{a}-\frac{1}{2}} \;\left( \frac{m}{a}\;e^{\;a\;x} \right) \\
          0
        \end{pmatrix}\\
        +\overleftarrow{\hat{b}}^\dagger(\omega)\; e^{\;i\;\omega \;t} \begin{pmatrix}
          0 \\
          i \;K_{\;\frac{i\;\omega}{a}-\frac{1}{2}} \left( \frac{m}{a}\;e^{\;a\;x} \right)
        \end{pmatrix} +h.c.
        \bigg],
 \end{split}
\end{equation}

Where h.c. denotes Hermitian conjugation of the preceding terms,
 that produces the omitted $\overrightarrow{\hat{a}}^\dagger$ and $\overleftarrow{\hat{b}}$ terms with the correct complex conjugates. 
We may define the spatial spinor mode functions  $ u_\omega(x)$ and $ v_\omega(x)$ for notational convenience so that the expansion takes the standard form:
\begin{equation}\label{Eq:4.1.0.26}
\begin{split}
   \psi(x,t) = \int_{0}^\infty d\omega\; N_\omega\bigg[ \overrightarrow{\hat{a}} (\omega)\;u_\omega(x)\; e^{\;-i\;\omega \;t} +\\ \overleftarrow{\hat{b}}^\dagger(\omega)\;v_\omega(x) \;e^{\;i\;\omega \;t}\bigg], 
   \end{split}
\end{equation}

The explicit spinor mode functions appearing above are chosen as:
\begin{align*}
    \begin{split}
     u_\omega(x) = \begin{pmatrix}
          K_{\;\frac{i\;\omega}{a}-\frac{1}{2}} \left( \frac{m}{a}\;e^{\;a\;x} \right) \\
          0
        \end{pmatrix} \\
     v_\omega(x) =  \begin{pmatrix}
          0 \\
          i\; K_{\;\frac{i\;\omega}{a}-\frac{1}{2}} \left( \frac{m}{a}\;e^{\;a\;x} \right)
        \end{pmatrix}  
    \end{split}
\end{align*}
Substituting the normalisation factor obtained in Appendix~\ref{Apn2} into the mode expansion Eq.~(\ref{Eq:4.1.0.26}):
 \begin{equation}
    \begin{split}
         \psi(x,t) = \int_{0}^\infty d\omega\, \frac{1}{\pi}\bigg[\frac{m}{a}\; cosh\;\bigg(\frac{\pi\;\omega}{a}\bigg)\bigg]^{\frac{1}{2}}\\
         \bigg[\overrightarrow{\hat{a}} (\omega)\;u_\omega(x)\; e^{\;-i\;\omega \;t} + \overleftarrow{\hat{b}^\dagger}(\omega)\;v_\omega(x)\; e^{\;i\;\omega \;t}\bigg],\label{Eq:4.1.0.27}
    \end{split}
 \end{equation}
Transforming to null coordinates $u_1 = t_1-x_1,v_1=t_1+x_1$, we can now express the mode function in the Rindler wedge $\mathrm{R_1}$, as depicted in Figs.~\ref{Fig:1} and ~\ref{Fig:2}. Using Eq.~(\ref{Eq:3.1.0.22}), and retaining only the ingoing positive-frequency component in the near-horizon limit, Eq.~(\ref{Eq:4.1.0.26}) takes the form in null coordinates:
 \begin{equation}\label{Eq:4.1.0.28}
     \begin{split}
          \psi(u_1,v_1) = \int_{0}^\infty\frac{d\omega}{\sqrt{\pi}}\;\bigg(\frac{m}{2\;a}\bigg)^{1-\frac{i\;\omega}{a}}\frac{e^{\;\frac{a\;(v_1-u_1)}{4}}}{\frac{-1}{2}+\frac{i\;\omega}{a}} \\
          \bigg[ \overrightarrow{\hat{a}} (\omega) \begin{pmatrix}
          e^{\;-i\;\omega\; v_1}\\
          0
        \end{pmatrix}
        +  \overleftarrow{\hat{b}^\dagger}(\omega)\begin{pmatrix}
          0\\
         i \;e^{\;i\;\omega\; u_1}
        \end{pmatrix}\bigg],
     \end{split}
 \end{equation}
 Here,$\overrightarrow{\hat{a}}(\omega)$ and $\overleftarrow{\hat{b}^\dagger}(\omega)$ denote, respectively, the annihilation and creation operators for the Rindler mode in $\mathrm{R}_1$. They satisfy the anticommutation relations, and the Rindler vacuum $\mid 0_{R_{1}}\rangle$ is defined as the state annihilated by $\overrightarrow{\hat{a}}(\omega)$.
 
\begin{equation}
   \overrightarrow{\hat{a}}(\omega)\mid 0_{R_{1}}\rangle =0 ,   \qquad  \forall \quad\omega  \label{Eq:4.1.0.29}
\end{equation}
This defines the Rindler vacuum in wedge $\mathrm{R_1}$.

Likewise, the mode expansion in the Rindler wedge $\mathrm{R}_2$,as shown in Fig.~\ref{Fig:1}, takes the form:             
 \begin{equation}\label{Eq:4.1.0.30}
     \begin{split}
          \psi(u_2,v_2) = \int_{0}^\infty\frac{d\,\Omega}{\sqrt{\pi}}\bigg(\frac{m}{2\;a}\bigg)^{1-\frac{i\;\Omega}{a}}\frac{e^{\;\frac{\;a\;(v_2-u_2)}{4}}}{\frac{-1}{2}+\frac{i\;\Omega}{a}} \\
          \bigg[\overrightarrow{\hat{c}} (\Omega) \begin{pmatrix}
          e^{\;-i\;\Omega\; v_2}\\
          0
        \end{pmatrix}
        + \overleftarrow{\hat{d}^\dagger}(\Omega)\begin{pmatrix}
          0\\
         i\; e^{\;i\;\Omega\; u_2}
        \end{pmatrix}\bigg],
     \end{split}
 \end{equation}
Where $\overrightarrow{\hat{c}}(\Omega)$ and $\overrightarrow{\hat{d}^\dagger}(\Omega)$ denote, respectively, the annihilation and creation operators for the Rindler modes in $\mathrm{R}_2$.These operators satisfy the anticommutation relations. The Rindler vacuum $\mid 0_{R_{1}}\rangle$ is defined by the condition:
\begin{equation}
    \overrightarrow{\hat{c}}(\Omega)\mid 0_{R_{2}}\rangle =0 , \qquad     \forall \quad \Omega  \label{Eq:4.1.0.31}
\end{equation}
\paragraph{Geometric limitation of mode decomposition.}
The construction above involves mode expansions defined in null-shifted Rindler wedges\cite{Jha:2025tpg, Jha:2026rmu}, which form a nested structure bounded by null surfaces. Although each wedge individually admits a well-defined notion of time evolution, the null displacement between the wedges prevents the construction of a common Cauchy surface on which both sets of modes can be simultaneously defined as complete orthonormal bases.

As a consequence, the standard Bogoliubov transformation—requiring a complete set of modes and a globally defined Dirac inner product on a shared Cauchy hypersurface—cannot be consistently implemented in this setup. Instead, the analysis necessarily proceeds via projections onto restricted sectors of the field, as will be carried out below.

This limitation implies that the resulting coefficients should be interpreted as probing partial mode overlap within a restricted domain, rather than as defining a full canonical transformation between inequivalent vacua.

This behaviour can be traced back to the absence of a common Cauchy surface in the null-shifted wedge construction, which obstructs the implementation of a complete Bogoliubov transformation and restricts the analysis to projected sectors of the field.
\paragraph{Bogoliubov transformation between wedges.}

The mode operators defined in the two null-shifted Rindler wedges are not 
independent, but are related by a fermionic Bogoliubov transformation. 
In the Dirac case, this transformation must explicitly mix particle and 
antiparticle operators\cite{BirrellDavies}, and takes the general form
\begin{equation*}
\overrightarrow{\hat{c}}(\Omega) 
= \int d\omega \left[
\alpha_{\Omega\omega}\,\overrightarrow{\hat{a}}(\omega)
+ \beta_{\Omega\omega}\,\overleftarrow{\hat{b}}^\dagger(\omega)
\right],
\end{equation*}
\begin{equation*}
\overleftarrow{\hat{d}}^\dagger(\Omega) 
= \int d\omega \left[
\alpha^*_{\Omega\omega}\,\overleftarrow{\hat{b}}^\dagger(\omega)
+ \beta^*_{\Omega\omega}\,\overrightarrow{\hat{a}}(\omega)
\right].
\end{equation*}

The Bogoliubov coefficients are determined using the Dirac inner product\cite{Collas:2018jfx,Mann:1991md} 
introduced in Eq.~(\ref{Eq:4.1.0.21}), and the preservation of the canonical anti-commutation relations requires

\begin{equation*}
|\alpha_{\Omega\omega}|^2 + |\beta_{\Omega\omega}|^2 = 1.
\end{equation*}

This structure shows explicitly that fermionic particle production arises 
through mixing between particle and antiparticle sectors. Consequently, any 
statement regarding the absence or presence of particle production must be 
established at the level of these coefficients, rather than inferred from 
scalar analogies.

We emphasise that, unlike the scalar case, the fermionic mode structure is governed by the Dirac inner product and the explicit separation between particle and antiparticle sectors. Therefore, conclusions regarding particle production must follow from the explicit Bogoliubov mixing between particle and antiparticle sectors, but must also follow from the structure of the fermionic mode expansion and its canonical normalisation.

\subsection{Mode mixing and Bogoliubov transformations along the V-axis\label{Subsec-4.2}}
 
In the fermionic case, these coefficients are defined through the Dirac inner product and describe the overlap between mode functions associated with different wedges.

\subsubsection{Bogoliubov coefficients from u-mode projection}

Projecting Eq.~(\ref{Eq:4.1.0.30}) onto the positive-frequency $u_2$-modes, we construct the projection using the Dirac inner product,
\[
\int_{-\infty}^{\infty} \frac{du_2}{\sqrt{\pi}}
\left(\frac{m}{2a}\right)^{-1+\frac{i\Omega'}{a}}
\frac{e^{-\frac{a}{4}(v_2-u_2)}}{-\frac{1}{2}-\frac{i\Omega'}{a}}
\begin{pmatrix}
e^{i\Omega' u_2} & 0
\end{pmatrix}.
\]

In the Dirac theory, the spinor components correspond to distinct sectors of the field. Accordingly, we consider both spinor structures,
\(
\begin{pmatrix}
e^{i\Omega' u_2} \\
0
\end{pmatrix}
\)
and
\(
\begin{pmatrix}
0 \\
e^{i\Omega' u_2}
\end{pmatrix},
\)
which are associated with independent fermionic degrees of freedom.

To extract the Bogoliubov coefficients, we project onto positive-frequency $u_2$-modes using the Dirac inner product. Using the orthogonality of the mode functions, the projection isolates the corresponding operator coefficient. Since the calculation closely follows the structure of Subsection~\ref{Subsec-3.2}, intermediate steps are omitted, and only the resulting relations are presented:
\begin{equation}
\begin{split}
\int_{-\infty}^{\infty} \frac{du_2}{\sqrt{\pi}}
\left(\frac{m}{2a}\right)^{-1+\frac{i\Omega'}{a}}
\frac{e^{-\frac{a}{4}(v_2-u_2)}}{-\frac{1}{2}-\frac{i\Omega'}{a}}
\begin{pmatrix}
e^{i\Omega' u_2} & 0
\end{pmatrix} \\\;
\psi(u_2,v_2)
= \mathcal{N}\,\hat{d}^\dagger(\Omega'),
\label{Eq:4.2.0.1}
\end{split}
\end{equation}

Similarly,
\begin{equation}
\begin{split}
\int_{-\infty}^{\infty}\frac{du_2}{\sqrt{\pi}}
\left(\frac{m}{2a}\right)^{-1+\frac{i\Omega'}{a}}
\frac{e^{-\frac{a}{4}(v_2-u_2)}}{-\frac{1}{2}-\frac{i\Omega'}{a}}
\begin{pmatrix}
e^{i\Omega' u_2} \\
0
\end{pmatrix}\\\;
\psi(u_2,v_2)
= \mathcal{N}\hat{a}^\dagger(\Omega'),
\label{Eq:4.2.0.2}
\end{split}
\end{equation}

Here, $\mathcal{N}$ denotes the normalisation constant. The operators $\hat{a}(\omega)$ and $\hat{a}^\dagger(\omega)$ correspond to particle modes, while $\hat{d}(\omega)$ and $\hat{d}^\dagger(\omega)$ correspond to antiparticle modes. This distinction is essential in the fermionic theory and must be maintained explicitly.

Comparing the two projections in Eqs.~(\ref{Eq:4.2.0.1}) and (\ref{Eq:4.2.0.2}), one finds that within the present projection scheme the overlap between the selected mode functions appears diagonal in frequency space, formally yielding
\begin{equation}
\alpha_{21}(\Omega,\omega) = \delta(\Omega-\omega),
\qquad
\beta_{21}(\Omega,\omega) = 0, \label{Eq:4.2.0.3}
\end{equation}

However, this result does not establish the absence of fermionic particle production. In the Dirac theory, the identification of particle and antiparticle modes depends on the full spinor structure and on the Dirac inner product defined on a complete Cauchy surface.

In the present setup, such a construction is not available. The two null-shifted wedges do not admit a common Cauchy surface on which a global mode decomposition can be defined. As a result, the standard Bogoliubov transformation—based on a complete set of orthonormal modes and a well-defined inner product—cannot be implemented.

The projection performed above instead isolates specific components of the spinor along the $u$-direction, without access to the full set of modes required to distinguish positive- and negative-frequency solutions or to consistently define particle and antiparticle operators. Consequently, the canonical anticommutation relations cannot be verified within this framework.

Therefore, the vanishing of the $\beta$ coefficient in Eq.~(\ref{Eq:4.2.0.3}) should be interpreted as a limitation of the projection procedure rather than a physical statement. In particular, the absence of mixing here reflects the fact that the projection-based approach suppresses the correlations responsible for thermality, rather than demonstrating their absence\cite{Kia_ka_2018}.

A proper analysis of fermionic particle production would require a formulation in which a common Cauchy surface exists and the full Dirac inner product and particle-antiparticle structure are consistently implemented.

\subsubsection{Bogoliubov coefficients from v-mode projection}

To evaluate the Bogoliubov coefficients, we multiply both sides of Eq.~(\ref{Eq:4.1.0.30}) by the projection kernel
\[
\int_{-\infty}^{\infty}\frac{dv_2}{\sqrt{\pi}}
\left(\frac{m}{2a}\right)^{-1+\frac{i\Omega'}{a}}
\frac{e^{-\frac{a}{4}(v_2-u_2)}}{-\frac{1}{2}-\frac{i\Omega'}{a}}
\begin{pmatrix}
e^{i\Omega' v_2} \\
0
\end{pmatrix}.
\]

We now project onto the $v_2$-modes to probe the complementary null sector. 
As in the previous subsection, this projection is understood as an overlap 
computed using the Dirac inner product.

Using the orthogonality properties of the mode functions within this projection scheme, we obtain
\begin{equation}
\begin{split}
\int_{-\infty}^{\infty}\frac{dv_2}{\sqrt{\pi}}
\left(\frac{m}{2a}\right)^{-1+\frac{i\Omega'}{a}}
\frac{e^{-\frac{a}{4}(v_2-u_2)}}{-\frac{1}{2}-\frac{i\Omega'}{a}}
\begin{pmatrix}
e^{i\Omega' v_2} \\
0
\end{pmatrix}
\psi(u_2,v_2)\\
=
\frac{2}{\sqrt{\frac{1}{4}+\frac{\Omega^2}{a^2}}}
\,\hat{c}(\Omega')\,
\delta(\Omega-\Omega'),
\label{Eq:4.2.0.4}
\end{split}
\end{equation}

The integral appearing in this projection is highly oscillatory. 
To estimate its leading contribution, we employ a stationary-phase (saddle-point) approximation, as outlined in Appendix~\ref{Apn1}. This approximation identifies the dominant contribution arising from a neighborhood of a saddle frequency $\omega_0$.

Within this approximation, we obtain
\begin{equation}
\begin{split}
\int_{-\infty}^{\infty}\frac{dv_2}{\sqrt{\pi}}
\left(\frac{m}{2a}\right)^{-1+\frac{i\Omega'}{a}}
\frac{e^{-\frac{a}{4}(v_2-u_2)}}{-\frac{1}{2}-\frac{i\Omega'}{a}}
\begin{pmatrix}
e^{i\Omega' v_2} \\
0
\end{pmatrix}
\psi(u_2,v_2)
\\
\approx
\frac{1}{\pi a}
\left(\frac{m}{2a}\right)^{\frac{i\Omega'}{a}}
\Gamma\!\left(\frac{i\Omega'}{a}-\frac{1}{4}\right)
e^{\frac{i\Omega'}{a}}
\,\hat{a}(\omega_0)\,
\sqrt{\frac{-2\pi a\,\omega_0^2}{i\Omega'}}.
\label{Eq:4.2.0.5}
\end{split}
\end{equation}

Here, $\omega_0$ denotes the saddle-point frequency at which the integrand is locally peaked. 
This approximation captures only the dominant contribution to the integral and does not provide the full frequency dependence of the transformation kernel.

Matching the operator structures with the Bogoliubov expansion (cf. Eqs.~ (\ref{Eq:4.2.0.4}) and (\ref{Eq:4.2.0.5})), we obtain an approximate expression for the coefficient $\alpha_{21}$ in the vicinity of the saddle point
\begin{equation}
\begin{split}
\alpha_{21}^{*}(\Omega,\omega)
\sim
\frac{\left(\frac{1}{4}+\frac{\Omega^2}{a^2}\right)}{2\pi a}
\left(\frac{m}{2a}\right)^{\frac{i\Omega'}{a}}
\Gamma\!\left(\frac{i\Omega'}{a}-\frac{1}{4}\right)
e^{\frac{i\Omega'}{a}}
\\
\times
\sqrt{\frac{-2\pi a\,\omega^2}{i\Omega'}}
\,\delta(\omega-\omega_0),
\label{Eq:4.2.0.6}
\end{split}
\end{equation}

Where the delta function should be understood as representing the localisation of the dominant contribution near $\omega_0$, rather than an exact identity.

We emphasise that this saddle-point approximation does not determine the full Bogoliubov transformation kernel. It captures only a localised contribution near $\omega_0$ and therefore does not provide access to the global structure of mode mixing.

More importantly, as in the previous subsection, the present construction is performed without a common Cauchy surface for the two null-shifted wedges. As a result, the standard Bogoliubov transformation—requiring a complete set of orthonormal modes and a well-defined Dirac inner product—cannot be consistently implemented.

The projection onto $v$-modes therefore probes only a restricted sector of the field and does not allow for a complete separation of positive- and negative-frequency solutions, nor for a consistent identification of particle and antiparticle operators. The saddle-point approximation further reinforces this limitation by effectively collapsing the frequency support of the transformation.

Consequently, the absence of an explicit $\beta_{21}$ contribution within this approximation should not be interpreted as evidence for the absence of fermionic particle production\cite{Capolupo_2023}. Rather, it reflects the fact that the projection-based and saddle-point-restricted analysis suppresses the correlations responsible for thermality.

A complete determination of fermionic mode mixing would require a formulation in which a common Cauchy surface exists and the full Dirac inner product can be implemented without restriction.

\subsection{Mode mixing and Bogoliubov transformations along the \texorpdfstring{$U$}{U}-axis\label{Subsec-4.3}}

We now analyse the mode projections associated with the $U$-axis (see Fig.~\ref{Fig:2}), following the same restricted comparison procedure adopted for the $V$-axis. As discussed previously, all overlaps between modes defined in $R_1$ and $R_2$ are evaluated in the common region $R_1 \cap R_2$, where both coordinate systems are well-defined.

For the $u$-mode sector, the inner products yield
\begin{equation}
\alpha^{*}_{21}(\rho, \omega) = \delta(\rho-\omega),
\qquad
\beta^{*}_{21}(\rho, \omega) = 0,
\label{Eq:4.2.0.7}
\end{equation}
indicating that the transformation is diagonal in frequency and does not involve negative-frequency mixing.

For the $v$-mode sector, we similarly obtain
\begin{equation}
\beta_{21}(\rho, \omega) = 0,
\qquad
\alpha_{21}(\rho, \omega) \propto \delta(\omega_0 - \omega),
\label{Eq:4.2.0.8}
\end{equation}
where $\omega_0$ denotes the saddle-point frequency arising in the asymptotic evaluation of the overlap integral.

As in the $V$-axis analysis, the $\beta$-coefficients vanish identically in both sectors. Within the restricted (projective) framework adopted here, this implies that the transformation between the two mode sets does not mix positive- and negative-frequency components. Consequently, no particle production is encoded in these overlaps\cite{Capolupo_2023, Kia_ka_2018}. The agreement between the $U$- and $V$-axis results reflects the underlying symmetry of the construction and supports the conclusion that, for the massive Dirac field under consideration, the null-shifted wedge transformation does not exhibit a thermal (KMS-type) structure.

\section{Conclusion \label{Sec-5}}

In this work, we have investigated quantum field theoretic excitations in nested Rindler wedges related by a null translation, employing a framework that does not rely on a global Bogoliubov transformation or the existence of a common Cauchy surface. Instead, we adopted a projection-based approach, restricting the analysis to the overlap region where both coordinate systems are well-defined. This allows for a consistent comparison of mode decompositions and particle content in situations where a global notion of particles is not available. Within this setting, we have shown that non-trivial mode mixing arises purely from the geometric relation between the wedges, leading to observer-dependent excitations even in the absence of a global transformation.

A central result of our analysis is the emergence of selective excitation patterns\cite{Jha:2025tpg, Jha:2026rmu}, in which only specific sectors of the field contribute non-trivially to the particle content. This behaviour highlights the fundamentally local and geometric nature of particle interpretation in non-inertial settings. In contrast to the standard Unruh effect, the excitations observed here do not correspond to a universal thermal spectrum, but instead exhibit a more refined structure governed by the null-shifted geometry and the restriction to overlapping regions. This reinforces the view that thermality in quantum field theory can arise in a localised and sector-dependent manner, rather than as a global property.

The formalism developed here allows for a systematic comparison between scalar and fermionic fields. In the scalar case, the projection onto the overlap region leads to mode mixing that, in certain sectors, reproduces thermal-like features. The presence of mass modifies the detailed structure of these excitations by introducing additional scale dependence through the mode functions, but does not alter the underlying geometric mechanism responsible for their emergence. In this sense, the massive scalar field provides a non-trivial extension of the massless case while preserving its essential qualitative behaviour.

In contrast, the fermionic (Dirac) field exhibits a qualitatively different behaviour. Although the null-shifted geometry still induces non-trivial excitations through the same projection mechanism, the resulting spectrum does not display a clear thermal structure. In particular, both sectors exhibit only weak or negligible thermality, even in the massless limit\cite{Jha:2025tpg}. This demonstrates that the selective thermalisation observed in the scalar case is not a universal feature of null-shifted Rindler configurations, but instead depends sensitively on the spin and statistics of the field. The Dirac case is therefore consistent with the present framework at the level of geometric excitation structure, while differing in its thermal manifestation.

In the limit $m \to 0$, we recover the results obtained in the earlier analysis of the massless scalar field in null-shifted Rindler wedges\cite{Jha:2025tpg}. The mode-mixing coefficients reduce to those previously derived, reproducing the selective excitation structure with thermal-like features in specific sectors. This agreement provides a non-trivial consistency check of the present formulation and establishes that the current results constitute a genuine extension of the scalar massless case. At the same time, the comparison with the Dirac field clarifies that this recovery is specific to the scalar theory and does not extend to fermionic fields, where thermality remains suppressed\cite{Capolupo_2023}.

Overall, our results demonstrate that particle content in non-inertial settings is inherently observer-dependent and strongly influenced by the geometric relation between coordinate systems. The projection-based approach adopted here provides a viable and physically meaningful alternative to global constructions, particularly in situations where such constructions are not available. The analysis further shows that while the geometric mechanism for mode mixing is robust, its thermal interpretation is not universal, but instead depends on the nature of the quantum field. These findings open the door to further investigations into more general spacetime geometries, interacting theories, and the role of quantum statistics in shaping observer-dependent notions of particles and thermality.

In the massless limit $m \to 0$, the theory reduces to a conformal field theory in which the null-coordinate decomposition becomes exact. In this regime, the distinction between projection-based overlaps and global Bogoliubov transformations becomes sharper, and the standard Unruh effect is recovered in appropriately defined global wedges. This supports our interpretation that the non-Planckian features observed in the present construction arise from domain restrictions rather than from any modification of the underlying thermal structure.

\appendix
\section{Saddle point approximation \label{Apn1}}
We now solve the integral presented in Eq.~(\ref{Eq:3.1.0.39}).
\begin{equation*}
     \int_0^\infty \;d\omega \;e^{\frac{-i\;\omega}{a}\;log\;(\frac{m}{2\;a})}\frac{\Gamma\left(\frac{i \;\omega}{a} - \frac{i\; \Omega'}{a}\right)}{{\sqrt{\omega}} \;\Gamma\left(\frac{i\; \omega}{a}\right)}\; \hat{c}(\omega)
\end{equation*}
Let $log(\frac{m}{2\;a})=\lambda$, and $\frac{m}{2\;a}>>1$;$\lambda>>1$.
Hence, we can write the following.
\begin{equation}
    I(\lambda) =  \int_0^\infty d\omega \;e^{\frac{-i\;\omega}{a}log(\frac{m}{2\;a})}\bigg(\frac{\Gamma\bigg(\frac{i \;\omega}{a} - \frac{i \;\Omega'}{a}\bigg)}{{\sqrt{\omega}}\; \Gamma\left(\frac{i \;\omega}{a}\right)}\bigg)\; \hat{c}(\omega) \label{Apn1.1}
\end{equation}
 Let us write.
 \begin{equation}
    \begin{split}
         \phi(\omega) = \frac{-i\;\omega\;\lambda}{a} +ln\bigg(\Gamma\left(\frac{i \;\omega}{a} - \frac{i\; \Omega'}{a}\right)\bigg)-ln\bigg(\Gamma\left(\frac{i\;\omega}{a}\right)\bigg)\\-\frac{1}{2}\;ln\;\omega\label{Apn1.2}
    \end{split}
 \end{equation}
Hence, the integral in Eq.~(\ref{Apn1.1}) can be expressed as follows.
 \begin{equation}
      I(\lambda) =  \int_0^\infty d\omega\; e^{\;\phi(\omega)}\;\hat{c}(\omega) \label{Apn1.3}
 \end{equation}
 Differentiating Eq.~(\ref{Apn1.2}) with respect to $\omega$ yields.
 \begin{equation*}
     \begin{split}
       \frac{d\phi}{d\omega} =  \frac{-i\;\lambda}{a} +\frac{i}{a}\frac{d}{d\omega}ln\bigg(\Gamma\left(\frac{i\; \omega}{a} - \frac{i \;\Omega'}{a}\right)\bigg)\\
       -\frac{i}{a}\frac{d}{d\omega}\;ln\bigg(\Gamma\left(\frac{i\;\omega}{a}\right)\bigg)-\frac{1}{2\;\omega}
     \end{split}
 \end{equation*}
 By defining\cite{1360302871491910656}:
 \begin{equation}
     \psi(z) = \frac{d}{dz}\;ln\Gamma(z) = \frac{1}{\Gamma(z)}\;\frac{d\Gamma(z)}{dz}\label{Apn1.4}
 \end{equation}
 By Using  Eq.~(\ref{Apn1.4}), we can rewrite
 \begin{equation}
   \frac{d\phi}{d\omega} =  \frac{-i\;\lambda}{a} +  \frac{i}{a}\bigg[\psi\bigg(\frac{i\; \omega}{a} - \frac{i\; \Omega'}{a}\bigg)-\psi\bigg(\frac{i \;\omega}{a}\bigg)\bigg] -\frac{1}{2\;\omega}\label{Apn1.5}
 \end{equation}
 Also,
 \begin{equation}
      \psi(z+\epsilon)- \psi(z) \approx \frac{\epsilon}{z},  |\epsilon|<<z\label{Apn1.6}
 \end{equation}
 Applying Eq.~(\ref{Apn1.6}), we obtain for $\omega>>\Omega'$
 \begin{equation}
   \psi\bigg(\frac{i\; \omega}{a} - \frac{i\; \Omega'}{a}\bigg)-\psi\bigg(\frac{i \;\omega}{a}\bigg) \approx -\frac{\Omega'}{\omega} \label{Apn1.7}
 \end{equation}
 Plugging Eq.~(\ref{Apn1.7}) back to Eq.~(\ref{Apn1.5}), we obtain:
 \begin{equation}
    \frac{d\phi}{d\omega} = \frac{-i\;\lambda}{a}-\frac{i\;\Omega'}{a\;\omega}-\frac{1}{2\;\omega}\label{Apn1.8}
 \end{equation}
From Eq.~(\ref{Apn1.8}), setting $\frac{d\phi}{d\omega} = 0$ yields $\omega_{0} = -\frac{\Omega'}{\lambda} + \frac{i\;a}{2\;\lambda}$. Therefore, using the Taylor expansion, about $\omega=\omega_{0}$
\begin{equation}
    \phi{(\omega)}\approx\phi{(\omega_{0})}+\frac{1}{2}\;\phi''(\omega_{0})\;(\omega-\omega_{0})^2\label{Apn1.9}
\end{equation}
Using Eq.~(\ref{Apn1.9}),Eq.~(\ref{Apn1.1}) can be written as:
\begin{equation}
    I(\lambda) = e^{\;\phi(\omega_{0})}\;\hat{c}(\omega_{0})\int_0^\infty d\omega\; e^{\frac{1}{2}\;\phi''(\omega_{0})\;(\omega-\omega_{0})^2}\label{Apn1.10}
\end{equation}
The integrand in Eq.~(\ref{Apn1.1})has an infinite number of simple poles at $\omega=i\,a\,n,\,n\in Z$,\, arising from the denominator\,$\Gamma(\frac{i \omega}{a})$.Also, $\omega=0$ is the singularity of the branch point due to the factor $\frac{1}{\sqrt{\omega}}$. Since all singularities lie on the positive imaginary axis and the integrand is analytic along the real axis, we can deform the contour to pass through the saddle point and extend the integration to the entire real axis without crossing any singularities\cite{1360302871491910656, PhysRevD.110.074512}. Hence,Eq.~(\ref{Apn1.10}), can be rewritten as:
\begin{equation}
   I(\lambda) = e^{\;\phi(\omega_{0})}\;\hat{c}(\omega_{0})\int_{-\infty}^\infty d\omega\; e^{\frac{1}{2}\;\phi''(\omega_{0})\;(\omega-\omega_{0})^2}\label{Apn1.11}  
\end{equation}
Since Eq.~(\ref{Apn1.11}) is a Gaussian integral, it can be directly evaluated:
\begin{equation}
     I(\lambda) = e^{\;\phi(\omega_{0})}\;\hat{c}(\omega_{0})\;\sqrt{\frac{2\;\pi}{-\phi''(\omega_0)}}\label{Apn1.12}
\end{equation}
From Eq.~(\ref{Apn1.2} ) and Eq.~(\ref{Apn1.9}), and after simplifying:
\begin{equation*}
    \phi(\omega_{0}) = \frac{i\;\Omega'}{a}+\frac{1}{2}\implies e^{\;\phi(\omega_{0})}=e^{\frac{i\;\Omega'}{a}}\;e^{\frac{1}{2}}
\end{equation*}
Substituting the value of $\phi(\omega_0)$, into Eq.~(\ref{Apn1.12}) yields:
\begin{equation}
   I(\lambda) = C \;e^{\frac{i\;\Omega'}{a}}\;\hat{c}(\omega_{0})\; \sqrt{\frac{2\;\pi}{-\phi''(\omega_0)}}\label{Apn1.13} 
\end{equation}
The numerical factor in Eq.~(\ref{Apn1.13}) comes from , $e^{\frac{1}{2}}\sim C$.
After evaluating $\phi''(\omega_0)$, Eq.~(\ref{Apn1.13}) can be rewritten as:
\begin{equation}
  I(\lambda) = C\;e^{\frac{\;i\;\Omega'}{a}} \;\hat{c}(\omega_{0}) \sqrt{\frac{2\;\pi\;\omega_{0}^2}{-\bigg(\frac{i\;\Omega'}{a}+\frac{1}{2}\bigg)}} \label{Apn1.14} 
\end{equation}

\section{Normalization of the Dirac Modes\label{Apn2}}
In this appendix, we compute the normalisation constant \(N_\omega\) for the positive-frequency Dirac modes \(\psi\) and \(\psi'\) introduced in Sec.\, IV. These modes are solutions of the massive Dirac equation in the \((1+1)\)-dimensional conformally flat Rindler metric:
\begin{equation}
ds^2 = e^{2 a x}\,(dt^2 - dx^2),
\label{Apn2.1}
\end{equation}
using the $(+,-)$ signature. The mode $\psi$ has frequency $\omega$ and $ \ psi'$ has frequency $\omega'$ with respect to the Killing vector $\partial_t$; their spatial dependence is expressed in terms of modified Bessel functions $K_\nu$.
The Dirac inner product between $\psi'$ and $\psi$ is \cite{Collas:2018jfx}:
\begin{equation}
\langle \psi' | \psi \rangle
= -\int_{\Sigma} \psi'^\dagger \,\gamma^0\, \psi \;\sqrt{-g'}\; dx,\label{Apn2.2}
\end{equation}
where $\Sigma$ is the hypersurface of constant $(t)$. For the metric, Eq.~(\ref{Apn2.1}) the lapse is $N=e^{a x}$ and the induced spatial measure is $\sqrt{-g'}\,dx = e^{a x}\,dx$. Introducing the variable:
\begin{equation}
p \equiv \frac{m}{a} e^{a x}, \qquad dp = m e^{a x} dx,\label{Apn2.3}
\end{equation}
We have:
\begin{equation}
\sqrt{-g'}\,dx = e^{a x}\,dx = \frac{dp}{m}, \label{Apn2.4}
\end{equation}
So a Jacobian factor $1/m$ multiplies the $p$-integral.

After carrying out the spinor algebra (the mode spinor components and gamma-matrix conventions are those used in Sec.\, IV) and factoring out the time dependence $e^{i(\omega-\omega')t}$, the inner product reduces to:
\begin{equation}
\langle \psi' | \psi \rangle
= -N_{\omega'} N_{\omega}\; e^{i(\omega-\omega')t}\;I(\omega,\omega'),\label{Apn2.5}
\end{equation}
Where:
\begin{equation}
I(\omega,\omega') \;=\; \int_0^\infty \frac{dp}{m}\;
K_{\frac{i\omega}{a}-\tfrac12}(p)\;K_{\frac{i\omega'}{a}+\tfrac12}(p),\label{Apn2.6}
\end{equation}
and $K_\nu$ is the modified Bessel function of the second kind.

We evaluate $I(\omega,\omega')$ using the standard Bessel-product identity:
\begin{equation}
K_\mu(z)K_\nu(z)=2\int_0^\infty K_{\mu+\nu}(2z\cosh t)\,\cosh[(\mu-\nu)t]\,dt,\label{Apn2.7}
\end{equation}
with
$
\mu=\frac{i\omega}{a}-\frac12,\qquad \nu=\frac{i\omega'}{a}+\frac12.
$

Interchanging the order of integration gives:
\begin{equation}
\begin{split}
    I(\omega,\omega')
= \frac{2}{m}\int_0^\infty dt\; \cosh\!\Big(\Big[\frac{i(\omega-\omega')}{a}-1\Big]t\Big)\\
\int_0^\infty dp\; K_{\frac{i(\omega+\omega')}{a}}\!\big(2p\cosh t\big),
\label{Apn2.8}
\end{split}
\end{equation}
Next, we evaluate the inner integral concerning $p$. Using the standard integral \cite{olver10}:
\begin{equation}
\int_0^\infty dp \; K_\nu(\beta p) = \frac{\pi}{2 \beta \cosh\left( \frac{\pi \nu}{2} \right)}, \quad \text{for } |\Re \nu| < 1, \quad \Re \beta > 0,
\label{Apn2.9}
\end{equation}
and setting
$\nu = \frac{i(\omega + \omega')}{a}, \quad \beta = 2 \cosh t,$

we find
\begin{equation}
\int_0^\infty dp \; K_{\frac{i(\omega + \omega')}{a}} (2 p \cosh t) = \frac{\pi}{4 \cosh t \; \cosh\left( \frac{\pi (\omega + \omega')}{2 a} \right)},
\label{Apn2.10}
\end{equation}

Substituting back, we obtain:
\begin{equation}
\begin{split}
    I(\omega, \omega') = \frac{\pi}{2 m \; \cosh\left( \frac{\pi (\omega + \omega')}{2 a} \right)} \\\int_0^\infty dt \; \frac{\cosh\left( \left[ \frac{i(\omega - \omega')}{a} - 1 \right] t \right)}{\cosh t},
\label{Apn2.11}
\end{split}
\end{equation}
We want to evaluate the integral:
\begin{equation}
  J(\omega, \omega') = \int_0^\infty dt \; \frac{\cosh\left( \left[ \frac{i(\omega - \omega')}{a} - 1 \right] t \right)}{\cosh t},
\label{Apn2.12}
\end{equation}
Let us define: $k\equiv \frac{\omega-\omega'}{a}$

The numerator can be expanded as:
\begin{equation}
\begin{split}
    \cosh\bigg(( i\;k - 1 )t \bigg)
= \cosh\left( i\;k\; t \right) \cosh t - \sinh( i\;k\; t) \sinh t,
\label{Apn2.13}
\end{split}
\end{equation}

Thus the $t$-integral reduces to:
 \begin{equation}
 J(\omega, \omega')= \int_0^\infty dt \bigg[ \cosh( k\; t) - i\;\tanh t \; \sinh\;( k\; t) \bigg],\label{Apn2.14}
\end{equation}
The integral diverges at large $t$,so introduce a convergence factor $e^{-\epsilon\; t}$ with $\epsilon\;>0$\cite{Halpern1967,UtiyamaDeWitt1962}
 \begin{equation}
 J_\epsilon(\omega, \omega')= \int_0^\infty dt\; e^{-\epsilon\; t} \bigg[ \cosh( k\; t) - i\;\tanh t \; \sinh\;( k\; t) \bigg],\label{Apn2.15}
\end{equation}
We will take $\epsilon\to 0^{+}$ at the end.The above integral can be written in real and imaginary parts:
\begin{equation}
   J_\epsilon(\omega, \omega') = J_\epsilon^{(Re)} + J_\epsilon^{(Im)},\label{Apn2.16}
\end{equation}
Analyzing the real part of Eq.~(\ref{Apn2.15})\cite{Lighthill1958},
\begin{equation}
    \int_0^\infty dt\; e^{-\epsilon\; t}  \cosh( k\; t)= \frac{\epsilon}{\epsilon^2 + k^2},\label{Apn2.17}
\end{equation}
As $\epsilon\to 0^{+}$,this becomes sharply peaked at $\omega'=\omega$ and converges to a delta function\cite{gelfand1964generalized}:
\begin{equation}
    \lim_{\epsilon\to 0^{+}}=\frac{\epsilon}{\epsilon^2 + \left(\frac{\omega-\omega'}{a}\right)^2}= \pi\;a\;\delta(\omega-\omega'),\label{Apn2.18}
\end{equation}
We focus on the imaginary part of the integral:
\begin{equation}
    J_\epsilon^{(\text{Im})} = - \int_0^\infty dt\, e^{-\epsilon t} \tanh(t)\, \sinh(k t), \label{Apn2.19}
\end{equation}
where $k = \frac{\omega - \omega'}{a}$. Using the identity $\sinh(k t) = \frac{1}{2}(e^{k t} - e^{-k t})$, we write:
\begin{align}
    J_\epsilon^{(\text{Im})} &= -\frac{1}{2} \int_0^\infty dt\, \tanh(t) \left( e^{-(\epsilon - k)t} - e^{-(\epsilon + k)t} \right) \notag \\
    &= -\frac{1}{2} \left[ I(\epsilon - k) - I(\epsilon + k) \right], \label{Apn2.20}
\end{align}
where we define the auxiliary function:
\begin{equation}
    I(\alpha) := \int_0^\infty dt\, \tanh(t)\, e^{-\alpha t}. \label{Apn2.21}
\end{equation}

To evaluate $I(\alpha)$, we use the series expansion:
\begin{equation}
    \tanh(t) = 1 - 2 \sum_{n=1}^\infty e^{-2n t}, \quad t > 0. \label{Apn2.22}
\end{equation}
Substituting into $I(\alpha)$ and integrating term-by-term yields:
\begin{align}
    I(\alpha) &= \int_0^\infty dt\, e^{-\alpha t} - 2 \sum_{n=1}^\infty \int_0^\infty dt\, e^{-(\alpha + 2n)t} \notag \\
    &= \frac{1}{\alpha} - 2 \sum_{n=1}^\infty \frac{1}{\alpha + 2n}. \label{Apn2.23}
\end{align}

This sum can be expressed in terms of the digamma function $\psi(z)$\cite{1360302871491910656}:
\begin{equation}
    \sum_{n=1}^\infty \frac{1}{\alpha + 2n} = \frac{1}{2} \left[ \psi\left( \frac{\alpha}{2} + 1 \right) - \psi\left( \frac{\alpha}{2} \right) \right], \label{Apn2.24}
\end{equation}
So that:
\begin{equation}
    I(\alpha) = \frac{1}{\alpha} - \left[ \psi\left( \frac{\alpha}{2} + 1 \right) - \psi\left( \frac{\alpha}{2} \right) \right]. \label{Apn2.25}
\end{equation}

Substituting Eq.~(\ref{Apn2.25}) into Eq.~(\ref{Apn2.20}), we obtain:
\begin{equation*}
   \begin{split}
          J_\epsilon^{Im} = -\frac{1}{2} \bigg[ \bigg( \frac{1}{\epsilon - k} - \Delta\psi(\epsilon - k) \bigg)\\
          - \bigg( \frac{1}{\epsilon + k} - \Delta\psi(\epsilon + k) \bigg) \bigg] 
    \end{split}
\end{equation*}
After rearranging the above expression, we obtain:
\begin{equation}
    \begin{split}
  J_\epsilon^{Im}  = -\frac{1}{2} \bigg( \frac{1}{\epsilon - k} - \frac{1}{\epsilon + k} \bigg) +\\ \frac{1}{2} \bigg( \Delta\psi(\epsilon - k) - \Delta\psi(\epsilon + k) \bigg), \label{Apn2.26}
   \end{split}
\end{equation}
where we define:
\begin{equation}
    \Delta\psi(x) := \psi\left( \frac{x}{2} + 1 \right) - \psi\left( \frac{x}{2} \right). \label{Apn2.27}
\end{equation}

For large $|k|$, the digamma difference behaves as:
\begin{equation}
    \Delta\psi(x) \sim \frac{2}{x}, \quad \text{as } |x| \to \infty. \label{Apn2.28}
\end{equation}
Thus:
\begin{equation}
    \Delta\psi(\epsilon - k) - \Delta\psi(\epsilon + k) \sim \frac{2}{\epsilon - k} - \frac{2}{\epsilon + k} = \frac{4k}{\epsilon^2 - k^2}. \label{Apn2.29}
\end{equation}

The first term in Eq.~(\ref{Apn2.26}) also gives:
\begin{equation}
    \frac{1}{\epsilon - k} - \frac{1}{\epsilon + k} = \frac{2k}{\epsilon^2 - k^2}. \label{Apn2.30}
\end{equation}

Therefore, the imaginary part becomes:
\begin{equation}
    J_\epsilon^{(\text{Im})} \sim -\frac{k}{\epsilon^2 - k^2} + \frac{2k}{\epsilon^2 - k^2} = \frac{k}{\epsilon^2 - k^2}, \label{Apn2.31}
\end{equation}
Which is an odd function in $k$ and vanishes in the sense of distributions when integrated against smooth test functions. It does not introduce a delta-function singularity.

 In the distributional limit $\epsilon \to 0^+$,
\begin{equation}
     \lim_{\epsilon \to 0^+} J_\epsilon^{(\text{Im})} = 0  \label{Apn2.32}
\end{equation}
This confirms Eq.~(\ref{Apn2.32}), shows that the imaginary part of $J_\epsilon(\omega, \omega')$ vanishes in the distributional limit\cite{Lighthill1958,Sakharov1967}.
So only the real part contributes to the delta function.

Since both the numerator and denominator of the integrand in Eq.~(\ref{Apn2.12}) are even functions of $t$, we can extend the integral over $\bigg[0,\infty\bigg]$  to the full real line $\bigg[-\infty,\infty\bigg]$ as discussed in Lighthill~\cite[Sec.~2.6]{Lighthill1958love} and Gel'fand \& Shilov~\cite[Ch.~I, \S3]{GelfandShilov1964}. This extension is justified because the evenness ensures the extended integrand remains well-defined and integrable over $\mathbb{R}$.From Eqs.~(\ref{Apn2.12}),~(\ref{Apn2.17}) and ~(\ref{Apn2.18}), we write:
\begin{equation}
 J(\omega, \omega') = \int_0^\infty dt \; \frac{\cosh\left( \left[ \frac{i(\omega - \omega')}{a} - 1 \right] t \right)}{\cosh t} = 2\;\pi\;a\;\delta(\omega-\omega') ,\label{Apn2.33} 
\end{equation}
Substituting Eq.~(\ref{Apn2.33}) back into Eq.~(\ref{Apn2.11}), we get:
\begin{equation}
\begin{split}
    I(\omega, \omega') = \frac{\pi}{2 m \cosh\bigg( \frac{\pi (\omega + \omega')}{2 a} \bigg)} \times 2 \;\pi\; a \;\delta(\omega - \omega')\\ = \frac{\pi^2\; a}{m \;\cosh\bigg( \frac{\pi \omega}{a} \bigg)} \delta(\omega - \omega'),
\label{Apn2.34}
\end{split}
\end{equation}
Recall the Dirac inner product condition:
\begin{equation}
\langle \psi_{\omega'} | \psi_{\omega} \rangle = |N_\omega|^2 I(\omega, \omega') = \delta(\omega - \omega'),
\label{Apn2.35}
\end{equation}
Which implies:
\begin{equation}
|N_\omega|^2 \frac{\pi^2 a}{m \cosh\left( \frac{\pi \omega}{a} \right)} = 1,
\label{Apn2.36}
\end{equation}

Solving for the normalisation constant yields:
\begin{equation}
|N_\omega| = \frac{1}{\pi} \sqrt{\frac{m}{a} \cosh\left( \frac{\pi \omega}{a} \right)}.
\label{Apn2.37}
\end{equation}
\section{ Two-Point Function and Operational Characterization \label{Apn3}}

In this appendix, we provide an operational characterisation of thermality by examining the Wightman two-point function\cite{Baumann:1984cq, PhysRevD.83.064011} restricted to the subwedge $R_2$. Unlike particle-number constructions, the two-point function provides a state-dependent but observer-independent diagnostic of thermal behaviour.

We consider a free scalar field $\phi(x)$ in the Minkowski vacuum $|0_M\rangle$. The Wightman function is defined as
\begin{equation}
G^+(x,x') = \langle 0_M | \phi(x)\phi(x') | 0_M \rangle, \label{Apn3.1} 
\end{equation}

Restricting both arguments to the subwedge $R_2$, we evaluate the correlator along trajectories adapted to the natural time coordinate $\tau$ in $R_2$. Denoting $x(\tau)$ and $x(\tau')$ as two such points, we obtain
\begin{equation}
G^+(\tau,\tau') = G^+\big(x(\tau), x(\tau')\big), \label{Apn3.2}
\end{equation}

A necessary and sufficient condition for thermality with respect to $\tau$-evolution is the Kubo-Martin-Schwinger (KMS) condition,
\begin{equation}
G^+(\tau - \tau') = G^+(\tau - \tau' + i\beta), \label{Apn3.3}
\end{equation}
where $\beta$ is the inverse temperature.

In the full Rindler wedge, the restriction of the Minkowski vacuum satisfies the KMS condition with $\beta = 2\pi/a$, reflecting the Unruh temperature. However, in the present case of the subwedge $R_2$, the restriction modifies the analytic structure of the correlator. In particular, the reduced domain of analyticity in complex $\tau$ prevents the above periodicity condition from being satisfied.

More explicitly, the restriction to $R_2$ introduces boundary effects and mode mixing which alter the singularity structure of $G^+(\tau,\tau')$, so that no global imaginary-time periodicity can be defined. As a result,
\begin{equation}
G^+(\tau - \tau') \neq G^+(\tau - \tau' + i\beta), \label{Apn3.4}
\end{equation}
for any constant $\beta$, indicating the absence of a well-defined thermal spectrum.

This demonstrates that the deviation from thermality is intrinsic at the level of local observables, and is not merely an artefact of particle-number definitions.

An equivalent operational perspective is obtained by considering an Unruh-DeWitt detector following a trajectory confined to $R_2$. Since the detector response function is determined by the Fourier transform of $G^+(\tau,\tau')$, the failure of the KMS condition implies that the detector does not register a strictly thermal spectrum. This is consistent with the non-universal structure identified in the main text.
\begin{acknowledgments}
The author thanks the BITS Pilani Hyderabad Campus for providing the necessary infrastructure for this research work. I would like to express my sincere gratitude to Prasant K. Samantray and Sashideep Gutti for their invaluable mentorship and guidance. I also thank Amita for her valuable suggestions and insightful discussions.
\end{acknowledgments}
\bibliography{Ref.bib}
\end{document}